\documentclass[preprint]{aastex}
\def\sqd{{deg$^{2}$}}
\def\etal{{\it et al.}}
\def\kms{km~s$^{-1}$}
\def\msun{$M_\odot$}
\def\be{\begin{equation}}
\def\ee{\end{equation}}

\begin{document}

\title{The Arecibo Legacy Fast ALFA Survey:\\ 
     II. Results of Precursor Observations 
%     {\it Version 3.1 \hskip 8pt 15May09}
}

\author {Riccardo Giovanelli\altaffilmark{1}, Martha P. Haynes\altaffilmark{1}, 
Brian R. Kent\altaffilmark{1}, Philip Perillat\altaffilmark{2}, 
Barbara Catinella\altaffilmark{2}, G. Lyle Hoffman\altaffilmark{3}, 
Emmanuel Momjian\altaffilmark{2}, Jessica L. Rosenberg\altaffilmark{4}, 
Amelie Saintonge\altaffilmark{1}, Kristine Spekkens\altaffilmark{1}, 
Sabrina Stierwalt\altaffilmark{1}, Noah Brosch\altaffilmark{5}, 
Karen L. Masters\altaffilmark{1}, Christopher M. Springob\altaffilmark{1}, 
Igor D. Karachentsev\altaffilmark{6}, Valentina
E. Karachentseva\altaffilmark{7}, Rebecca A. Koopmann\altaffilmark{8}, 
Erik Muller\altaffilmark{9}, Wim van Driel\altaffilmark{10},
Liese van Zee\altaffilmark{11}}

\altaffiltext{1}{Center for Radiophysics and Space Research
and National Astronomy and Ionosphere Center, 
Cornell University, Ithaca, NY 14853. {\it e--mail:} riccardo@astro.cornell.edu,
haynes@astro.cornell.edu, bkent@astro.cornell.edu, amelie@astro.cornell.edu,
spekkens@astro.cornell.edu, sabrina@astro.cornell.edu, masters@astro.cornell.edu, 
springob@astro.cornell.edu}

\altaffiltext{2}{Arecibo Observatory, National Astronomy and Ionosphere Center,
Arecibo, PR 00612. {\it e--mail:} phil@naic.edu, bcatinel@naic.edu, emomjian@naic.edu}

\altaffiltext{3}{Hugel Science Center, Lafayette College, Easton, PA 18042.
{\it e--mail:} hoffmang@lafayette.edu}

\altaffiltext{4}{Harvard--Smithsonian Center for Astrophysics, 60 Garden St. MS 65, 
Cambridge MA 02138--1516. {\it e--mail:} jlrosenberg@cfa.harvard.edu}

\altaffiltext{5}{The Wise Observatory \& The School of Physics and Astronomy, 
Raymond \& Beverly Sackler Faculty of Exact Sciences, Tel Aviv University, Israel.
{\it e--mail:} noah@wise.tau.ac.il}

\altaffiltext{6}{Special Astrophysical Observatory, Russian Academy of Sciences, 
Niszhnij Arkhyz 369167, Zelencukskaya, KChR, Russia. {\it e--mail}: ikar@sao.ru}

\altaffiltext{7}{Dept. of Astronomy \& Space Science, Kyiv University, Kyiv 
252017, Ukraine. {\it e--mail:} vkarach@observ.univ.kiev.ua} 

\altaffiltext{8}{Dept. of Physics \& Astronomy, Union College, Schenectady, NY 12308.
{\it e--mail:} koopmanr@union.edu}

\altaffiltext{9}{ATNF, CSIRO, PO Box 76, Epping, NSW 1710, Australia; {\it e--mail:} 
erik.muller@atnf.csiro.au} 

\altaffiltext{10}{Observatoire de Meudon, 5 Place Jules Janssen, 92195 Meudon, France.
{\it e--mail:} wim.vandriel@obspm.fr}

\altaffiltext{11}{Astronomy Dept., Indiana University, Bloomington, IN  47405. 
vanzee@astro.indiana.edu}

\begin{abstract}
In preparation for the full Arecibo Legacy Fast ALFA extragalactic HI survey, precursor
observations were carried out in Aug--Sep 2004 with the 7-beam Arecibo L-band
feed array (ALFA) receiver system and the WAPP spectral processors. While these observations
were geared mainly at testing and debugging survey strategy, hardware and software, approximately
48 hours of telescope time yielded science--quality data. The efficiency of 
system usage (allowing for minor malfunctions and the impact of radio frequency
interference) during that time was 75\%. From those observations, an
initial list of 730 tentative detections of varying degree of
reliability was extracted. Ninety--eight high signal-to-noise candidates
were deemed to be {\it bona fide} HI line
detections. To test our ability to discriminate cosmic
signals from RFI and noise, 165 candidates ranging in reliability
likelihood were re--observed with the single
beam L--band wide system at Arecibo in Jan--Feb 2005. Of those, 41\% were confirmed as real.
We present the results of both the ALFA and single beam observations for the sample 
of 166 confirmed HI sources, as well as our assessment of their optical counterparts. 
Of the 166 sources, 62 coincide with previously known HI sources,
while optical redshifts were available for an additional 18 galaxies; thus, 52\% 
of the redshifts reported here were previously unknown. Of the 166 HI detections,
115 are identified with previously cataloged galaxies, of either known or 
unknown redshift, leaving 51 objects identified for the first
time. Because of the higher sensitivity of the Arecibo system,
fewer than 10\% of the 166 HI sources would have been
detected by a HIPASS--like survey of the same region. Three of the objects have HI
masses less than $10^7$ \msun. The full ALFALFA survey which commenced
in February 2005 should detect more than 100 times as many objects of
similarly low HI mass over the next 5 years.
\end{abstract}


\section {Introduction}

The upgrade of the surface of the Arecibo antenna in the mid-1970's
initiated a new era of extragalactic 21 cm HI line studies which
exploited the big dish's collecting area and rapid progress in ancillary
instrumentation (low noise amplifiers; broadband, flexible multi-bit
spectrometers). Throughout the last 30 years, Arecibo observations have 
contributed to studies of large scale structure and the local 
peculiar velocity field, the dark matter content of galaxies and the 
impact of environment on galaxy evolution. The number of galaxies for 
which HI has been detected at Arecibo exceeds 10000, the vast majority 
of which have been studied via pointed observations of their optical 
counterparts using Arecibo's single beam feed systems. HI blind 
surveys conducted at Arecibo during the period of the upgrade by 
Zwaan \etal ~(1997; the Arecibo HI Strip Survey: AHISS) and by 
Rosenberg \& Schneider (2000; the Arecibo Dual Beam Survey: ADBS)
demonstrated the potential for discovery by the combination of Arecibo's
high sensitivity and relatively small beam size, but were limited by
the small areal coverage feasible with a single beam, surveying only
65 and 430 \sqd, respectively.

In addition to greatly increasing the sensitivity and reducing the
frequency dependence of the Arecibo telescope, the Gregorian optical
system installed in the 1990's delivered the possibility of extending
Arecibo even further, through the installation of multi-feed arrays
(Kildal \etal ~1993).
Similar systems at Parkes and Jodrell Bank have enabled the first
full view of the extragalactic HI sky through the
Parkes Multi-Beam HI Survey (HIPASS; Barnes \etal ~2001) and the 
HI Jodrell All-Sky Survey (HIJASS; Lang \etal ~2003). The HIPASS catalog
(HICAT; Zwaan \etal ~2004, Meyer \etal ~2004) includes 4315 detected HI sources
over an area of 21341 \sqd. The advantages of Arecibo in terms of gain,
angular and spectral resolution promise a significantly greater detected
population, with much lower confusion probability and a deeper sampled
volume, despite its more limited sky coverage 
($-2^\circ < Decl. < +38^\circ$).

Most recently, the Arecibo telescope has been outfitted
with a seven--beam 21~cm system dubbed the Arecibo L--Band Feed Array (ALFA).
The combination of ALFA with the very large collecting area
provided by the 305~m primary mirror now enables the undertaking of large
area, high sensitivity surveys of the 21~cm sky to probe a variety of 
astronomical phenomena. To nurture the best use of ALFA for such large scale 
surveys, the National Astronomy and Ionosphere
Center (NAIC) has fostered the formation of broad consortia, focused
on separate science topics ranging from pulsar searches, to Galactic studies of continuum
emission, the HI line and radio recombination lines, to extragalactic 
HI and OH line research. For extragalactic studies, the availability of ALFA
renews the Arecibo legacy of high sensitivity,
probing regimes untouched by other surveys and addressing
fundamental cosmological questions (the number density, distribution
and nature of low mass halos) and issues of galaxy formation and
evolution (sizes of HI disks, history of tidal interactions and mergers,
low z absorber cross section, origin of dwarf galaxies, nature
of high velocity clouds). 

The current authors constitute the initial team interested in exploring
possibilities for conducting a blind HI survey of the nearby
universe and thus came together to answer NAIC's call for initial survey
precursor observations that would explore, develop and test optimal observing
strategies, calibration schemes and data reduction pipelines. We have 
recently initiated a very wide area (7000 \sqd) blind HI survey: the 
Arecibo Legacy Fast ALFA (ALFALFA) survey. The ALFALFA program
science goals and strategy are described in a companion paper
(Giovanelli \etal ~2005; Paper I). The precursor observations
were carried out during the commissioning phase of ALFA and the data
thus obtained do not lend themselves to a statistically accurate astronomical
analysis. However, they provide an interesting new set of data and
they allow for `lower--limit' estimates of the potential of the ALFALFA 
survey, which we explore. We thus present those results and discuss how 
they aided us in adopting the final ALFALFA survey strategy.

In Section \ref{observations}, we briefly discuss the observations made using
the ALFA system as well as follow-up observations made with a single-pixel
feed to test detection reliability. In Section \ref{results}, we present the
parameters of HI detections. Their probable optical counterparts are
identified in Section \ref{optid}. Section \ref{parmqual} and Section \ref{sensitivity}
discuss the quality of inferred parameters and survey sensitivity.
The final section summarizes the experience gleaned from the precursor
observations discussed here in the context of the full ALFALFA survey
as it is now being undertaken.

\section{Observations}\label{observations}

As part of the campaign to commission the ALFA instrument, NAIC issued
a call for proposals in early 2004 for ``precursor'' programs that would
engage the potential user community in the system commissioning. The
observations reported here were carried out with the ALFA system
in August and September 2004 under a ``shared-risk'' policy, allowing
us to gain experience with the instrumentation, to test methodology
and to develop software necessary for data management and processing.
During this period, information about the ALFA system performance 
was accumulated and documented at the NAIC ALFA website\footnote
{\it http://alfa.naic.edu/memos}; a number of the memos posted there 
were contributed by members of this observing team. In this section,
we briefly describe the configuration of the ALFA system and the
observing mode adopted for the precursor observations. More details
on the system can be found in Paper I. It should be noted that the system
and our approach have evolved since the precursor observations were
undertaken, partly in response to the experience gained through them.

\subsection{ALFA Observations}\label{ALFAobs}

Given the ``shared risk'' character of these early ALFA observations,
a large fraction of the
telescope time was used to test and optimize data-taking procedures and
to debug and calibrate the new
system. Data in ``survey mode'' were obtained for a total of 47.8 hours, 
albeit a fraction of those were affected by several problems. Among the
problems encountered during this ``shakedown'' period were
the loss of one of the polarization channels on the central pixel (beam 0); 
strong, permanent and internally generated radio frequency
interference (RFI) on the other polarization channel of beam 0;
transient, internally generated RFI with drifting frequency over all 
beam/polarization channels (the `wandering birdie'); strong external RFI,
mixing selectively at different frequencies in each backend board; 
the effects of dynamically evolving data taking software. Most of those 
problems were resolved or largely attenuated towards the end of or after 
our precursor run. The efficiency of telescope usage over the 47.8 hours 
of our observations in survey mode during this observing session was $\simeq 75$\%.
One important consequence of the problems encountered was the unevenness
of the sky coverage on small scales and an impact on the statistical
homogeneity of the data sets.

The gain of the seven beams of ALFA ranges between 8.5 and 11 K Jy$^{-1}$,
and the system temperatures range between 26 and 30 K on cold sky (including
atmosphere and Cosmic Microwave Background emission).

All the survey mode observations were made with the telescope stationed 
at azimuth=180$^\circ$, thus enabling observations of the sky north of 
the Arecibo zenith. The feed array (see Paper I for a description) was 
rotated to an angle of 19$^\circ$ from its nominal rest alignment, 
yielding a configuration in which the sky footprints of its seven 
individual beams sweep tracks on the sky equally spaced in Declination,
as the Earth rotates. The spacing of the beams in this configuration is 
$\sim$2.1\arcmin ~(126\arcsec), slightly worse than Nyquist, as the
beams have elliptical half power full size of $3.3$\arcmin$\times 3.8$\arcmin. 
Drift scans of 900 seconds duration were taken sequentially, with about 
3--4 seconds of dead time between one scan and the next. The duration 
of each drift was dictated by the need to obtain individual data files 
of manageable size and to minimize data loss in case of equipment 
malfunction. A successive set of drifts was obtained with minimal 
interruption. It should be noted that under the constraints of telescope 
control during the 2004 precursor period, the beams followed tracks of 
constant declination at the current epoch, although initial positioning 
was commanded to J2000 Declination. ALFALFA observations in 2005 allow 
for periodic adjustments to maintain a constant J2000 Declination.

%Data taking interruption was 
%necessary in order to (a) obtain individual data files of manageable size, 
%(b) reset the Declination to a constant 2000.0 epoch (as a stationary 
%telescope would track current Declination), (c) minimize data loss in case 
%of equipment malfunction.  

The backend spectrometer system consisted of a set of four WAPPs (Wideband
Arecibo Pulsar Processor), which allowed instantaneous autocorrelation and
sampling of 16 time series, each yielding a 100 MHz wide spectrum of 4096 
channels. Fourteen of those were used to process and record the two 
independent polarization channels of each of the seven beams of ALFA. The 
remaining two recorded redundant data and the resulting data buffers were 
subsequently used by us to record the median of all the 7 beams, as 
a tool for RFI monitoring. The spectral resolution of 25 kHz corresponds 
to $R=\lambda/\Delta \lambda\simeq 57,000$, or 5.3 \kms, at the 1420 MHz
rest frequency of the HI line. All spectra were centered at a frequency of 
1385 MHz, in the Observatory reference frame, i.e. no Doppler tracking was 
implemented to allow for the varying component of the Earth's motion with respect
to a cosmic reference frame. A heliocentric correction was computed and
applied to the data during the off--line processing stages, using an IDL 
routine developed by one of us (PP) and C. Heiles. The effective 
spectral coverage of the data was restricted to the interval between 
1342 to 1428 MHz, due to 
the drop in sensitivity of the backend near the band edges. Further losses 
of spectral coverage resulted from RFI, as described in the next section.
The range of radial velocities corresponding to that frequency interval, 
for the HI line, is $cz = [17500$ to $-1600]$ \kms. That interval thus includes 
the HI emission of the Milky Way and of perigalactic High Velocity Clouds,
as well as the rest frequency of several radio recombination lines. 

Data were recorded every second; thus a 900 second drift consists of 
900 records, each of sixteen, 4096 spectral channel spectra (two 
polarizations for each of the 7 ALFA beams plus one spare pair). A drift 
is made up of $900\times 8\times 2\times 4096$ spectral samples, each 
4 bytes long, plus header information, adding up to about 0.25 GByte.
This configuration differs from the ALFALFA survey currently under way,
for which drifts are 600 seconds long. The change was made in order to 
better accommodate the gain calibration scheme, which evolved from that 
adopted during the precursor observations.

\begin{figure}
%\figurenum{1}
%\plotone{./figs/pos.ps}
\plotone{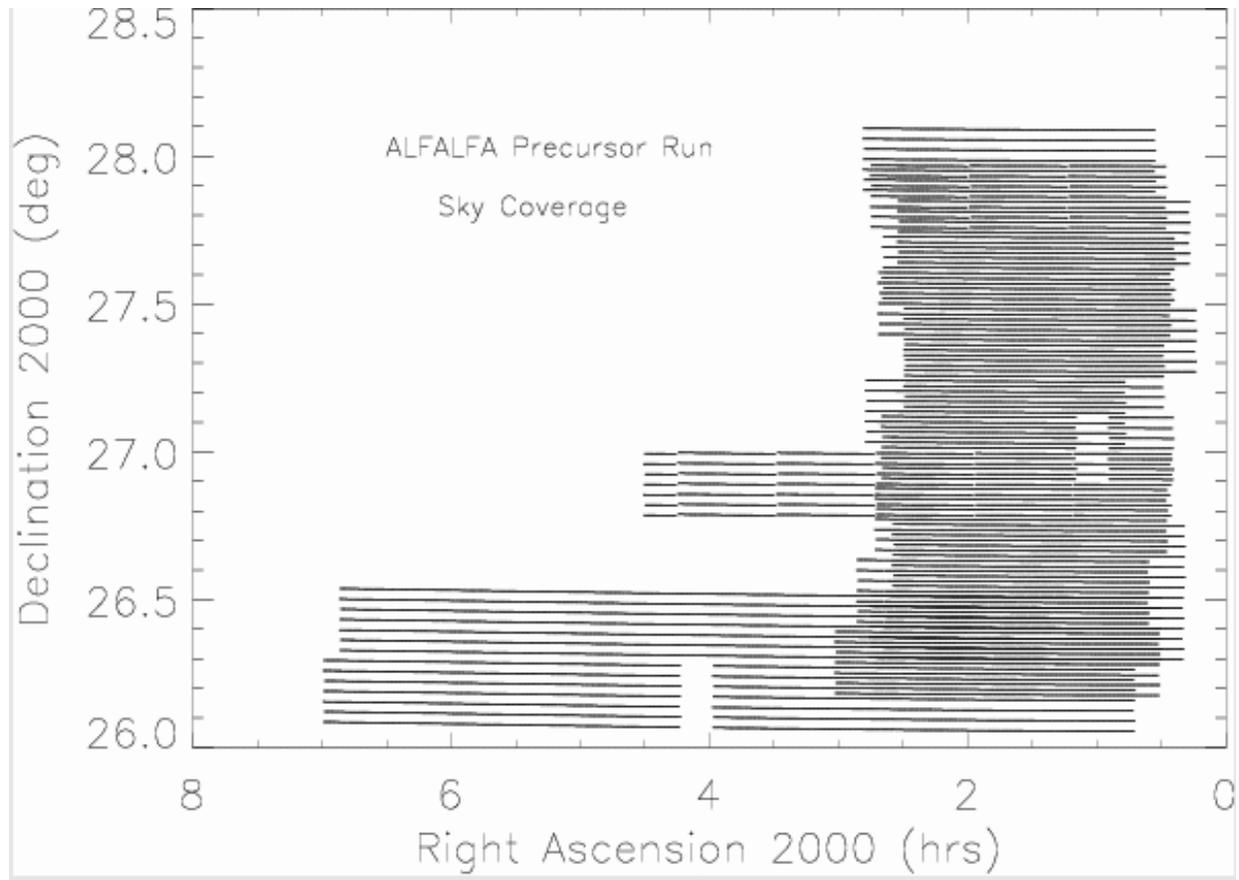}
\caption{Region sampled by the ALFALFA precursor run discussed in this
paper. Individual beam tracks for each drift are outlined by sequences of 
dots at constant Declination. Tracks in ``single--pass'' regions are
spaced 2.1\arcmin ~in Declination, while those in `two--pass' regions are
spaced 1.05\arcmin ~in Declination. Many gaps in local coverage are
present on small scales. For the full ALFALFA survey, sampling will be
more homogeneous.}
\label{positions}
\end{figure}

The sky coverage of the precursor run observations is shown in Figure
\ref{positions}. Part of the region, extending near 100$^\circ$ in
R.A. but confined to a swath only 0.5$^\circ$ in Declination. This region
was covered in ``single--pass'' mode, i.e. only one set of drift scans 
swept any part of the sky, with the Declination tracks 
of each beam separated from one another by $\sim$2.1\arcmin. A
$30^\circ \times 1.8^\circ$ region was (almost) covered in ``two--pass'' 
mode, whereby a second set of drift scans was obtained, sweeping over
the same region of sky during a separate observing session. During the
second pass, the center beam pointing shifted 7.3\arcmin
~in Declination. In this case, as evident in Figure \ref{positions},
the Declination sampling is twice as dense (1.05\arcmin), and the 
scalloping that arises from the fact that the central beam has higher
gain is smoothed somewhat. The second mode is representative
of the strategy adopted for the full ALFALFA survey (see Paper I).

%FIGURE 1

A noise calibration diode which increased the 
system power by about a third was fired for one second every 275 seconds 
during data taking. This turned out to be an inconvenient choice, for the
duration of the calibration was unstable, yielding spurious calibration
results. In latter parts of the run, the calibration diode was fired
for 2 seconds, assuring that the noise diode was ``on'' for at least
one full record. In that instance, 4 seconds (records) of data were lost 
and interpolated across during data processing. The stability of the noise
calibration was checked by monitoring the polarization ratios of the
Galactic HI line, which falls within the observed bandwidth, and by
monitoring the ratios of that line's intensity among the various beams,
which stochastically should remain constant in a stable system (Giovanelli
\etal ~ALFALFA memo\footnote
{\it http://www.astro.cornell.edu/$\sim$haynes/pre204/docs/memo040920.pdf}). 
Drifts in the calibration values of up to 4\% 
were observed, over the period of about one month of the
precursor run observations. The final flux density scale on which the
results reported here are based is estimated to be accurate to within 4\%. 
Analysis of the continuum data for sources within the region surveyed
is currently under way and is expected, when completed, to allow us to 
tighten the calibration of the flux density scale of the ALFALFA
dataset to within 1--2\%.

\subsection {The Radio Frequency Interference Environment}\label{rfi}

Several known sources of RFI polluted the observed spectral band, as well 
as several unexpected ones. The most prominent among those is the FAA 
radar at 1350 MHz, which rendered useless an interval of variable width, 
averaging approximately 150 spectral channels (nearly 4 MHz). Our
observations are thus ``blind'' to HI sources redshifted to $cz$ between
about 15,200 and 16,000 \kms. RFI
features related to the FAA radar also appeared at 1380 MHz about 90\%
of the time, affecting some 30 spectral channels, and with lower impact
and intensity at other frequencies as well. Downlink transmission by the
Global Positioning System (GPS) constellation of satellites affected data between
1378 and 1384 MHz for irregular intervals of approximately 100 seconds, every few 
hours. Other sources of RFI external to the Observatory occasionally affected
a large and variable number of frequencies throughout the run. The estimated
reduction in the usage efficiency of the 86 MHz surveyed band (after exclusion 
of the band skirts) due to external sources of RFI is about 10\%.

Several internal sources of RFI were identified during the precursor run.
Most insidious was a feature of drifting frequency and relatively low intensity,
which affected the spectral region between 1385 and 1420 MHz. This ``wandering
birdie'' occurred erratically throughout all our precursor run and was
locally generated. Fortunately, its source was finally identified and 
corrected in early 2005. At least 5\% 
of our precursor data were irrecoverably marred by this feature. Other RFI 
features were found to be localized within a single WAPP. One polarization channel 
of the central beam of ALFA was malfunctioning and thus unavailable during most of 
the precursor observations, while the other polarization channel of the same beam
exhibited a strong RFI feature some 10 MHz wide centered near 1384 MHz.
 
The overall reduction of efficiency of usage of the 86 MHz surveyed spectral band,
due to the combined effect of external and internal sources of RFI and partial
(i.e. affecting only minor parts of the equipment but maintaining most capabilities
intact) hardware malfunction, is estimated to have been 25\%,
over the 47.8 hours of telescope time in survey data taking
mode. Because several of these RFI sources have been subsequently
eliminated, efficiency in the ALFALFA dataset will be considerably higher.
This expectation is being confirmed by the early runs of the survey.

The inspection of data in a ``global'' framework, i.e. as maps rather than
assemblies of individual spectra, is an absolute necessity in gauging the
likelihood that any signal be of cosmic origin rather than noise or RFI.
We developed an algorithm that maintains a dynamic memory of RFI features,
flagging spectral regions of locally enhanced RFI activity and ``unflagging''
them as RFI becomes statistically quiescent. While the technique does not
guarantee total reliability in the assessment of the nature of any given
signal, it helps exclude the vast majority of spurious ones quite effectively.
Further development of this approach will be a necessity in the analysis of
the ALFALFA survey data in order to maximize its efficiency and reliability.

\subsection{Confirmation Observations} \label{LBWobs}

As the signal-to-noise of candidate detections decreases, their
contamination by spurious noise ``signals'' in a survey such as 
ALFALFA quickly rises. Detection algorithms must thus be optimized 
to discriminate real from spurious signals as well as possible.
In early February 2005, 14 hours of telescope time were scheduled 
with the Arecibo single pixel L--band wide (LBW) feed system 
to test the results of detection reliability simulations (see Paper I)
through confirmation of ALFA detection candidates. Observations were 
scheduled in the late afternoon and early evening. This time of day 
is notoriously bad for broad--band spectral line observing due to the 
rapidly varying pattern of standing waves originating from solar 
radiation. Observing time was available in small slots between 
previously scheduled programs. Rather than waiting for the LST 
period associated with the precursor ALFA observations to shift 
to a more advantageous part of the diurnal cycle, i.e. the second 
half of the year, we used the available segments of telescope time, 
adopting measures to minimize the impact of standing wave variations. 
The usage of the time was thus not optimal, but we could observe 165 
detection candidates, spanning a range of reliability likelihood, in 
a timely manner.

The gain of LBW is 11 K Jy$^{-1}$ near 1385 MHz and its system temperature 
on cold sky is 27 K, with small variations about these values resulting 
from the telescope configuration. All detections were confirmed by LBW
observations at zenith angles of less than 15$^\circ$, for which
gain and temperature variations are within a few percent.

The LBW observations were made with the same spectrometer backend used
for the ALFA observations, covering a spectral band of 100 MHz, centered
near 1385 MHz. The early part of the LBW observations were made with 8192 
spectral channels, while the latter part of the run adopted an observing 
mode change to 4096 spectral channels (to circumvent a software bug), 
which respectively yielded channel 
separations of about 2.5 and 5 \kms. All data were smoothed to the same 
spectral resolution of 10 \kms, before measurement of detected features. 

Integration times for ON source observations varied between 1 and 3 minutes.
They were both preceded and followed by OFF source observations of comparable
duration, so that only one third of the integration time was on source.
The quality of the spectral baselines remained poor, albeit far
superior to those obtained with the traditional, one--sided ON--OFF
pattern commonly used for extragalactic, spectral line pointed observations.
Evidence of the impact of standing waves on this data is conspicuous in
our spectra. Nonetheless, confirmation of 68 candidate ALFA detections 
was possible. 

\section{Results}\label{results}

In addition to providing an opportunity to shakedown the hardware and
software associated with the new multi-beam system, the precursor
observations also produced a spectral dataset to verify ALFA's true
efficacy for extragalactic science. To process the ALFA spectral line
dataset, we have adopted the IDL processing environment and have made 
use of the suite of general purpose analysis tools available at the Arecibo
Observatory as well as new ones developed by us specifically for 
ALFALFA. The raw spectral data were converted from the FITS
format delivered by the data acquisition software to the form of IDL 
structures adopting the format defined in previous years at NAIC by
one of us (PP). While some details of the data processing refer only
to the present precursor dataset, given that system changes have taken place
since the commissioning phase, the basic approach to data analysis 
follows that discussed for the full ALFALFA survey as presented in
Paper I.

In this section, we describe separately the HI detections extracted from the ALFA 
drift observations and from the targeted LBW follow-up. Table \ref{stats}
summarizes the results' statistics, which are described below.
Spectra, images of optical counterparts and tabulated information on these
data can be access through the web directly
\footnote{\it http://egg.astro.cornell.edu/precursor}, 
or through the ALFALFA website\footnote{\it http://egg.astro.cornell.edu/alfalfa}.

\begin{deluxetable}{lr}
\tablewidth{0pt}
\tabletypesize{\scriptsize}
\tablecaption{Precursor Run Statistics \label{stats}}
\tablehead{
\colhead{Description}  & \colhead{Number}}
\startdata
ALFA Candidate Detections		&730 \\
Candidates reobserved with LBW		&165 \\
Detections confirmed by LBW		& 68 \\
ALFA internally corroborated		& 98 \\
 \\
Total confirmed Detections		&166 \\
 \\
N with very likely optical counterparts	&162 \\
N with previous HI detections		& 62 \\
N with previously known redshifts	& 80 \\
N with previously cataloged photometry	&115 \\
\enddata
\end{deluxetable}

\subsection{ALFA Detections}

Individual drift scans were processed through several distinct stages. The first 
consisted of bandpass subtraction, calibration and baselining; this
step also produced continuum time series for each beam. The spectral line 
data output of this stage is in the form of baselined arrays of spectra, 
ready for individual inspection as position--velocity (time--frequency) 
images. The construction of 3D maps, resulting 
from the combination of drift tracks at different Declinations,
was not possible with effectiveness with the precursor run data,
because of the numerous gaps in the data streams and deficiencies in the data 
quality. The data were thus visually inspected by one of us (RG) 
--- one drift track at a time --- 
after spatial smoothing to an effective integration time of $t_s=14$ sec, 
which yielded position-velocity maps with a median rms noise of 
$3.5\times (res/10)^{-1/2}$ mJy, as described in Paper I, where $res$ is
the spectral smoothing window in \kms. The data of single drift maps were 
smoothed at various resolution levels, before inspection. After signal 
identification, features identified in more than one beam track were
accumulated using a Gaussian weighting function of 2\arcmin ~half power kernel,
smoothed to a resolution of 16 \kms and measured to obtain velocities, 
widths and flux integrals, to a resolution varying between 11 and 16 \kms. 
The cross--referencing of the candidate detections among contiguous beam
tracks was made after signal identification, obtaining corroborating 
evidence for each candidate source. While this approach improved the final 
signal--to--noise ratio, it did {\it not} play a role in the initial 
identification of detection candidates. An automated matched
filter signal extraction algorithm which operates in the Fourier domain
is currently being applied to the data; visual and automated source
detections are being compared to calibrate the threshold of the algorithm.
Its results will be presented in a later report (Saintonge \etal ~in preparation).

An initial list of 730 candidate detections was obtained through the 
process described above; these ranged in quality from highly probable
detections to very low signal-to-noise candidates, the majority of
which are not likely to be real. The identification of such a large
list at this stage was made specifically for the purpose of testing
signal verification techniques. Of the 730 initial candidates, 98
were deemed to be reliable detections to a high level of confidence,
either because their existence was previously known or because the
corroborating evidence of detection in several beams was strong and
of high signal--to--noise.
They were not reobserved. Of the remaining less certain detections,
165 candidates were reobserved in the 
confirmation run in early February 2005, with the single pixel 
LBW feed at Arecibo, as reported in the next section. Priority was
given to those deemed most likely to be real, but a range of
candidates were included in the followup observations. Sixty--eight of the
reobserved candidate detections were confirmed, for a confirmation 
rate of 41\%. The remaining 467 candidate sources have not been 
reobserved and will not be discussed further in this preliminary
report. These statistics are summarized in Table 1.

Whenever possible, a region of typically 100 square arcminutes 
around each of the candidate sources within the precursor dataset 
was inspected and a map of the total HI emission in that region 
was produced, yielding a centroid position and a total HI line flux 
integral. Because of the uneven sky coverage of the precursor run, 
many of the regions surrounding candidate detections were only sparsely 
sampled, as noted in the comments to the tabulated data below and in the
large variations in the resulting sensitivities. As described before,
for each candidate detection a composite spectrum was produced and 
a centroid extracted, after combining data from all spectra within 
several arcminutes radius from the centroid position, with each spectrum 
weighed according to a Gaussian kernel which was equivalent to smoothing 
the data from the $3.3$\arcmin~$\times 3.8$\arcmin ~telescope beam size 
to a circular beam of $4$\arcmin ~full width at half power. The rms noise 
as measured in this coadded spectrum is tabulated as described below, after 
spectral smoothing of between 11 and 16 \kms ~resolution. The HI line flux 
integral, center velocity and spectral width were measured on each composite 
spectrum; they are listed in Table \ref{alfadet}. Whenever the flux integral 
obtained from the map yielded significant evidence for emission extended 
in comparison with the beam size, that value is also reported in Table 
\ref{alfadet}. In the vast majority, the confirmed detections are weak. 
The possible impact of sidelobe contamination on each source was checked
by inspection of the vicinity of each candidate detection. 

For the few large, highly extended sources, such as HI 014729.9+271958 
(IC 1727) and HI 014753.9+272555 (NGC 672), for which a detailed study of the
HI extent and characteristics is warranted, we defer to a later report.
Spectral line profiles of each detected source, after coaddition and
smoothing as described above, can be accessed 
via the web through the aforementioned sites.

Table \ref{alfadet} contains the parameters derived from the ALFA HI data, namely:
\begin{itemize}
\item Col. 1: source name, composed of the qualifier 'HI' and the
R.A. and Dec. (epoch J2000.0) of the HI centroid.
\item Col. 2: estimates of the errors in the measurement of the centroid position,
              respectively in seconds of time for the R.A. ($\epsilon_\alpha$) and 
              in seconds of arc for the Declination ($\epsilon_\delta$).
\item Col. 3: center heliocentric velocity $cz$ of the feature and
              estimate of the measurement error on $cz$, $\epsilon_{cz}$, in \kms; 
              when two non zero numbers are listed for $\epsilon_{cz}$, the second 
              is an estimate of the systematic component of the error, 
              while the first is a statistical error which depends on the
              shape of the signal and the signal--to--noise. The systematic
              error results from the observer's subjective choice of the 
              signal boundaries: when those are clear, the systematic error
              is nil; when strong uncertainty exists in making
              that choice, a visual guess is made by the observer of the
              potential amplitude of the uncertainty. The total error on
              $cz$ is the sum in quadrature of the two components of the error.
\item Col. 4: velocity width of the spectral feature, in \kms, measured at 50\% 
              level of intensity of two peaks on each side of the signal and its
              associated estimate of error, $\epsilon_{w}$. Note that the 
              measurement error on the width is effectively a fractional error,
              so that $W (\epsilon_w)= 100 (50)$ should be read as $100^{+50}_{-33}$.
              Widths are corrected for instrumental smoothing and relativistic 
              broadening, i.e. the tabulated values are
              $(W_{measured}^2 -res^2)^{-1/2}/(1+z)$, where $res$ is the spectral
	      resolution in \kms. No disk inclination 
              nor turbulent motion correction is applied to the widths.
\item Col. 5: rms of the composite HI spectrum, in mJy, measured `locally', i.e.
              over a baseline region within 10 MHz of the detected feature rather
              than over the full spectrum.  
\item Col. 6: Flux integral of the spectral feature measured on the composite
              spectrum $F_c$, in Jy \kms, and its associated statistical error,
              $\epsilon_f$, compounded with the overall uncertainty on the flux 
              calibration. This value is an underestimate of the true flux if the
              source is extended in comparison with the beam size.
\item Col. 7: In cases for which the ALFA data indicate significant angular extent
              of the HI source, an estimate of the total flux of the source derived
              from a map, $F_m$ is given. Note that for the majority of sources
              the signal--to--noise ratio is not sufficient to obtain a reliable
              map.
\item Col. 8: A letter ``L'' indicates that the ALFA detection is corroborated by 
              observations with the LBW single--pixel feed.
              The results of those observations are given in Table \ref{LBWdet}.
              An asterisk in this column indicates that comments are available for 
              the given source.  
\end{itemize}

\subsection{LBW Detections}

As stated in Section \ref{LBWobs}, follow-up observations made with the 
single-pixel LBW system were conducted in several small blocks of 
telescope time under far from optimal observing conditions, to test 
the reliability of 
candidate detection of low signal-to-noise. The LBW observations 
were generally centered on the position of what was initially 
identified as the optical counterpart of the HI source. In some
cases, additional observations on alternative interpretations of 
that association  were made, to ascertain its reality. The results 
are presented in Table \ref{LBWdet}, which lists:
\begin{itemize}
\item Col. 1: source name, as given in Table \ref{alfadet}, with coordinates 
              referred to the centroid of the ALFA observations.
\item Cols. 2, 3: Right Ascension and Declination of the LBW observation, 
              epoch J2000.0.
\item Col. 4: center heliocentric velocity $cz$ of the feature, and
              estimate of the error on $cz$, $\epsilon_{cz}$ in \kms,
              defined as described for the analogous quantities in
              Table \ref{alfadet}.
\item Col. 5: velocity width of the spectral feature and related uncertainty, 
              in \kms, as described in Table \ref{alfadet}.
\item Col. 6: rms of the HI spectrum, in mJy, measured `locally', i.e.
              over a baseline region within 10 MHz of the detected feature rather
              than over the full spectrum. 
\item Col. 7: flux integral and related uncertainty, in Jy \kms.
\item Col. 8: an asterisk indicates that comments are available for the source.
\end{itemize}
Entries in Table \ref{LBWdet} are cross--referenced in Table \ref{alfadet} 
by the presence of an ``L'' in Col. 8 of that table. Spectra can be accessed
through the aforementioned websites.

\section {Identification of Optical Counterparts}\label{optid}

Fields of $10\arcmin\times 10\arcmin$ around each HI detection were inspected 
for optical counterparts,
using DSS2 via {\it Skyview}\footnote{{\it Skyview} was developed and 
maintained under NASA ADP Grant NAS5--32068 under the auspices of the High 
Energy Astrophysics Science Archive Research Center at the Goddard Space 
Flight Center Laboratory of NASA.}, NED\footnote{The NASA/IPAC Extragalactic
Database (NED) is operated by the Jet Propulsion Laboratory, California
Institute of Technology, under contract with the National Aeronautics and 
Space Administration.} and our privately maintained data base of extragalactic
sources (the ``AGC'', for ``Arecibo General Catalog''). Optical counterparts
are identified for all but 4 of the 166 confirmed HI sources. In a few cases, 
the identification of an optical counterpart is ambiguous, as noted below. 
In at least one of those cases (HI 062218.4+262631), the simplest explanation 
for the lack of an optical counterpart is that the source region is affected by 
high optical extinction, in the vicinity of the galactic plane. 
Table \ref{optparms} lists the positions of the proposed optical counterparts, as follows:

\begin{itemize}
\item Col. 1: source name, as in Table \ref{alfadet}.
\item Col. 2: catalog number in the AGC; AGC numbers smaller than 13000 coincide
              with those of the Uppsala General Catalog (Nilson 1973). 
\item Col. 3,4: Right Ascension and Declination (epoch J2000.0) of the proposed
              optical counterpart of the HI source. These coordinates are
              either obtained from previous tabulations or directly measured
              using the {\it Skyview} facility. Positional
              accuracies are of order 1\arcsec.
\item Col. 5: Source distance in Mpc, $D_{cmb}$, obtained from the radial velocity
              corrected to the Cosmic Microwave Background rest frame, and
              assuming pure Hubble flow with $H_\circ=70$ \kms ~Mpc$^{-1}$.
\item Col. 6: Source distance in Mpc, $D_{pec}$, obtained from the measured
              redshift and from the peculiar velocity field model of
              Tonry \etal ~(2000), for $H_\circ=70$ \kms ~Mpc$^{-1}$. In
              addition to the error on the measurement of $cz$ and the
              systematics in the flow model (see Masters 2005), a
              ``thermal'' velocity component of 187 \kms ~as assumed by Tonry
              et al. (2000) introduces an additional uncertainty of 2.7 Mpc
              on each distance estimate.
\item Col. 7: logarithm in base 10 of the HI mass $M_{HI}$, expressed in solar
              units via $M_{HI}=2.356\times 10^5 D^2 \times F$, where $D$
              is the distance in Mpc given in Col. 6
              and $F$ is the flux integral obtained from Col. 7 (or Col. 6 if
              Col. 7 is blank) of Table \ref{alfadet}. 
\item Col. 8: an asterisk indicates that comments are available for the source.
\end{itemize}

Notes associated with the objects listed in Tables \ref{alfadet}, \ref{LBWdet} and 
\ref{optparms} follow:

{\footnotesize
\noindent HI 001709.7+271616: sparse sampling; HI Dec. poorly constrained. 

\noindent HI 002115.8+262318: Marginal LBW confirmation due to standing 
                   waves of solar radiation.

\noindent HI 002312.5+272644: previous HI det at cz=3905. 

\noindent HI 002818.8+272136: previous HI det at cz=9608.

\noindent HI 002905.1+272913: beware: feature v. near transient RFI.

\noindent HI 003855.1+265753: previous HI det at cz=5192.

\noindent HI 004034.8+270239: 2MASS object at 004042.3+270243, but opt id more likely with larger,
                   uncatalogued lsb galaxy nearby.

\noindent HI 004256.8+271522: previous HI det at cz=5289.

\noindent HI 004357.1+260731: previous HI det at cz=4898. Checked for HI emission on gal. at 
                   004349.8+261113: negative. Optical id as given in table confirmed.

\noindent HI 004404.2+261243: marginal, but matches previous HI det at cz=10082.

\noindent HI 004411.5+265042: previous HI det at cz=5187; extended HI.

\noindent HI 004645.5+275553: ALFA detection superimposed on RFI; while easily identified
 		   on position-velocity map of drift, flux, width and centroid very
                   unreliable. LBW data cleaner, corroborated by previously measured
                   cz of 6893 for CGCG 500-088.

\noindent HI 004650.7+262846: previous HI det at cz=4954.

\noindent HI 004716.1+274854: previous HI det at cz=4742.

\noindent HI 004803.3+273717: marginal detection, corroborated by previous HI det at cz=4949.

\noindent HI 004816.9+262806: previous HI det at cz=5269.

\noindent HI 004834.5+274100: previous HI det at cz=5214.

\noindent HI 005009.5+271328: previous HI det at cz=4922.

\noindent HI 005305.3+272909: 2MASS object at 005310.1+272837, but  opt id more likely with larger,
                   uncatalogued lsb galaxy nearby.

\noindent HI 010158.0+263010: previous HI det at cz=10069.

\noindent HI 010225.7+263713: previous HI det at cz=4992.

\noindent HI 010318.1+264747: opt id is double system; HI profile shows evidence of blend.
                   2MASS object at 010321.4+264737, but  opt id more likely with larger,
                   uncatalogued lsb system nearby.

\noindent HI 011302.6+273832: Checked for HI emission on gal. at  011309.8+273925: negative.
                   Optical id as given in table confirmed.

\noindent HI 011443.5+270813: previous HI det at cz=3618.

\noindent HI 011743.9+270006: 2MASS object at 011737.0+265902, but  opt id more likely with larger,
                   uncatalogued lsb galaxy nearby.

\noindent HI 012242.5+265157: close pair of gals (Mark 355/6, CGCG 481-004) at (a) 012240.8+265206 
                   and (b) 012243.1+265200, previously measured cz: 9187, 9068;
                   optical id given as center of pair; HI centroid marginally favors
                   id with (b).

\noindent HI 012403.9+270300: previous HI det at cz=4917.

\noindent HI 012405.2+280433: sparse sampling; HI Dec. poorly constrained.

\noindent HI 012607.5+275810: merging pair; previous HI det at cz=4021.

\noindent HI 012944.1+272249: sev. opt candidates in field, most notably: (a) 012943.6+272340 and
                   (b) 012947.0+272220, merging pair; (b) is IRAS F01270+2706, with opt
                   cz of 12566, matching HI cz. Nonetheless, optical id remains ambiguous.

\noindent HI 013716.0+262605: previous HI det at cz=3872.

\noindent HI 014105.8+272007: checked for emission with LBW 1 beam off N,S,E,W; source not
                   significantly extended.

\noindent HI 014147.3+273159: marginal detection, corroborated by previous HI det at cz=10828.

\noindent HI 014214.9+262202: previous HI det at cz=359.

\noindent HI 014428.3+275522: previous HI det at cz=4037.

\noindent HI 014441.4+271707: previous HI det at cz=420.

\noindent HI 014455.3+272942: marginal det by ALFA.

\noindent HI 014640.9+264754: previous HI det at cz=361.

\noindent HI 014653.7+280448: marginal ALFA detection, corroborated by previously measured opt cz=3590.

\noindent HI 014724.3+275312: NGC 670; previous HI det at cz=3703.

\noindent HI 014729.9+271958: IC 1727, v. extended, blends with NGC 672; HI flux integral much
                   larger than tabulated central beam flux. Accurate flux integral
                   will require detailed study.

\noindent HI 014753.9+272555: NGC 672, v. extended, blends with IC 1727; HI flux integral much
                   larger than tabulated central beam flux. Accurate flux integral
                   will require detailed study. Primary distance of 7.9 Mpc available.

\noindent HI 014837.9+273259: close pair (CGCG 482-017), with (a) 014835.2+273253 and (b) 014835.3+273326,
                   opt. cz=10979; HI centroid very marginally favors id with (a).

\noindent HI 014915.3+274248: previous HI det at cz=10753.

\noindent HI 015011.6+271145: previous HI det at cz=3502.

\noindent HI 015013.0+273842: previous HI det at cz=3537.

\noindent HI 015439.8+271111: RFI excised from edge of feature in ALFA spectrum.

\noindent HI 015519.2+275645: KK16; previous HI det at cz=206. Primary distance of 4.7 Mpc available.

\noindent HI 015917.6+270027: previously measured cz=5309.

\noindent HI 015937.0+272555: previous HI det at cz=5267. Extended, but HI map corrupted, unable
                   to obtain total flux.

\noindent HI 015952.5+262407: opt counterpart is chain of merging units.

\noindent HI 020133.9+262914: merging pair; previous HI det at cz=5146.

\noindent HI 020144.4+263227: previous HI det at cz=5001.

\noindent HI 020248.2+263434: previous HI det at cz=14593.

\noindent HI 020304.8+271222: sev. starlike features in field, but no unambiguous optical 
                   counterpart identified. Different interpretations of
                   spectral extent of signal adopted in ALFA and LBW spectra. Marginal det.

\noindent HI 020329.8+273909: highest cz of detections list.

\noindent HI 020343.0+261608: previous HI det at cz=5014. Only 4' away from HI 020353.5+261719.
                   Peak fluxes suggest that sidelobes do not play role in detection.

\noindent HI 020353.5+261719: Only 4' separation from HI 020343.0+261608. Peak fluxes suggest
                   that sidelobes do not play role in detection.

\noindent HI 020430.7+275454: previous HI det at cz=4699.

\noindent HI 020626.5+270152: previous HI det at cz=4964.

\noindent HI 020902.1+273202: previous HI det at cz=9856.

\noindent HI 020913.9+264536: marginal detection, marginal confirmation with LBW.

\noindent HI 020954.1+273147: triplet of small units; opt id tentatively assigned to largest unit.

\noindent HI 021404.3+275302: UGC 1718 = NGC 855 at opt. cz=594, classified as Elliptical.
                   Small spiral at 021405.8+275034: LBW observation centered on
                   it yields much lower flux integral. Most likely
                   optical id is UGC 1718; however, classified as
                   elliptical, the galaxy is more likely a
                   star-forming dwarf with a central bar and some
                   young star clusters.

\noindent HI 022102.1+274615: two 2MASS galaxies within 2.5', but larger, uncatalogued object
                   is better match as opt. counterpart.

\noindent HI 022103.9+270204: marginal ALFA detection, marginal LBW confirmation.

\noindent HI 022224.8+262552: previously measured opt cz=11179.

\noindent HI 022335.8+271851: marginal detection corroborated by previous HI det at cz=10645.

\noindent HI 022340.2+270927: previous HI det at cz=10647.

\noindent HI 022348.9+272848: previous HI det at cz=10706.

\noindent HI 022355.8+270618: previous HI det at cz=5470.

\noindent HI 022405.6+263900: sev. small LSB features in field, most notably: (a) 022402.0+263958
                   and (b) 022408.9+263952; optical id ambiguous.

\noindent HI 022459.4+260314: marginal detection corroborated by previous HI det at cz=10080.

\noindent HI 022533.4+264458: previous HI det at cz=10348.

\noindent HI 022538.8+271709: previous HI det at cz=8984.

\noindent HI 022558.5+271607: 2MASS galaxy with measured opt. cz=10493; in group with NGC 916 at
                   022547.6+271432, cz=9614. Marginal HI parameters.

\noindent HI 022609.4+273549: previous HI det at cz=9980.

\noindent HI 022617.1+260750: peculiar (double) gal (CGCG 483-047) at opt. cz=10128. Marginal HI det.

\noindent HI 022620.2+271315: Very marginal ALFA detection, corroborated by previous HI det at cz=10357.

\noindent HI 022629.9+273937: previous HI det at cz=9710.

\noindent HI 022632.1+274941: previous HI det at cz=9621.

\noindent HI 022741.1+271328: marginal detection corroborated by previous HI det at cz=5123.

\noindent HI 022742.7+261406: previous HI det at cz= 9504.

\noindent HI 022745.2+263507: previous HI det at cz=9789.

\noindent HI 022745.2+263507: merging system, tidal features.

\noindent HI 022751.5+275429: previously measured opt cz=10468.

\noindent HI 022816.3+261854: previous HI det at cz=5237.

\noindent HI 023052.0+261047: weak det, corroborated by previous HI det at cz=13289.

\noindent HI 023103.8+274053: previous HI det at cz=4587.

\noindent HI 023137.5+261010: previous HI det at cz=10869.

\noindent HI 023851.0+275109: previous HI det at cz=4587.

\noindent HI 024416.4+260648: opt id is galaxy in compact group HCG 020, at anomalous opt cz=10561
                   (other galaxies in group near 14500 \kms). Marginal det.

\noindent HI 024600.5+280145: previous HI det at cz=7954.

\noindent HI 024609.7+270247: previous HI det at cz=5728.

\noindent HI 024752.7+270607: U2272 at 024804.5+270609, detected at end of drift, 12s in RA off 
                   opt pos; opt pos not covered; tabulated flux is lower limit; det certain,
                   but HI parms highly unreliable due to incomplete sampling of source region.

\noindent HI 025122.4+263459: previously measured cz=7485.

\noindent HI 033716.0+262357: large (30\arcsec) galaxy at (a) 033722.5+262505 and lesser one at 
                   (b) 033719.6+262255; due to exceptionally large HI centroid error, 
                   optical id tentatively assigned to (a), but remains ambiguous. Marginal det.

\noindent HI 040226.1+264950: previous HI det at cz=5639.

\noindent HI 040328.6+262149: previous HI det at cz=7064.

\noindent HI 041904.2+261210: previous HI det at cz=3751.

\noindent HI 042739.5+260545: high obscuration: near galactic plane; ambiguous opt id.

\noindent HI 060125.5+260524: Wein 163 at cz=5919; source region
sparsely sampled; Dec centroid poorly constrained.

\noindent HI 060628.3+262314: previously measured cz=2717.

\noindent HI 062218.4+262631: optically invisible; high obscuration: ZOA.

\noindent HI 063840.8+263006: previously measured cz=9819.

\noindent HI 065004.2+262342: previously measured cz=9509.
}

Of the 166 HI detections, 62 coincide with previously known HI sources. Redshifts 
had been previously measured for an additional 18 galaxies, so that 52\% 
of the redshifts reported here were previously unknown. Of the 166 objects
detected, optical identifications are made with 115 previously cataloged galaxies, 
while 51 are previously unreported objects. Images of the HI detections' 
optical counterparts can be accessed through the aforementioned websites.

\section{Quality of Inferred Parameters}\label{parmqual}

In the final definition of the ALFALFA survey design and strategy that
these precursor observations were meant to aid, evaluation of the useful
quantities derived from the spectral dataset provided quantitative, albeit
approximate, measures of survey strategy effectiveness. In this section, 
we use the precursor dataset to present a preliminary analysis of the 
parameters to be extracted from the ALFALFA survey dataset.

\subsection{HI Positions}

\begin{figure}[h]
%\figurenum{1}
% either psfig or epsscale+plotone will work below.
%\centerline{\psfig{figure=../rpe2_6.ps}}
%\epsscale{1.0}
%\plotone{./figs/pos_histo.ps}
\plotone{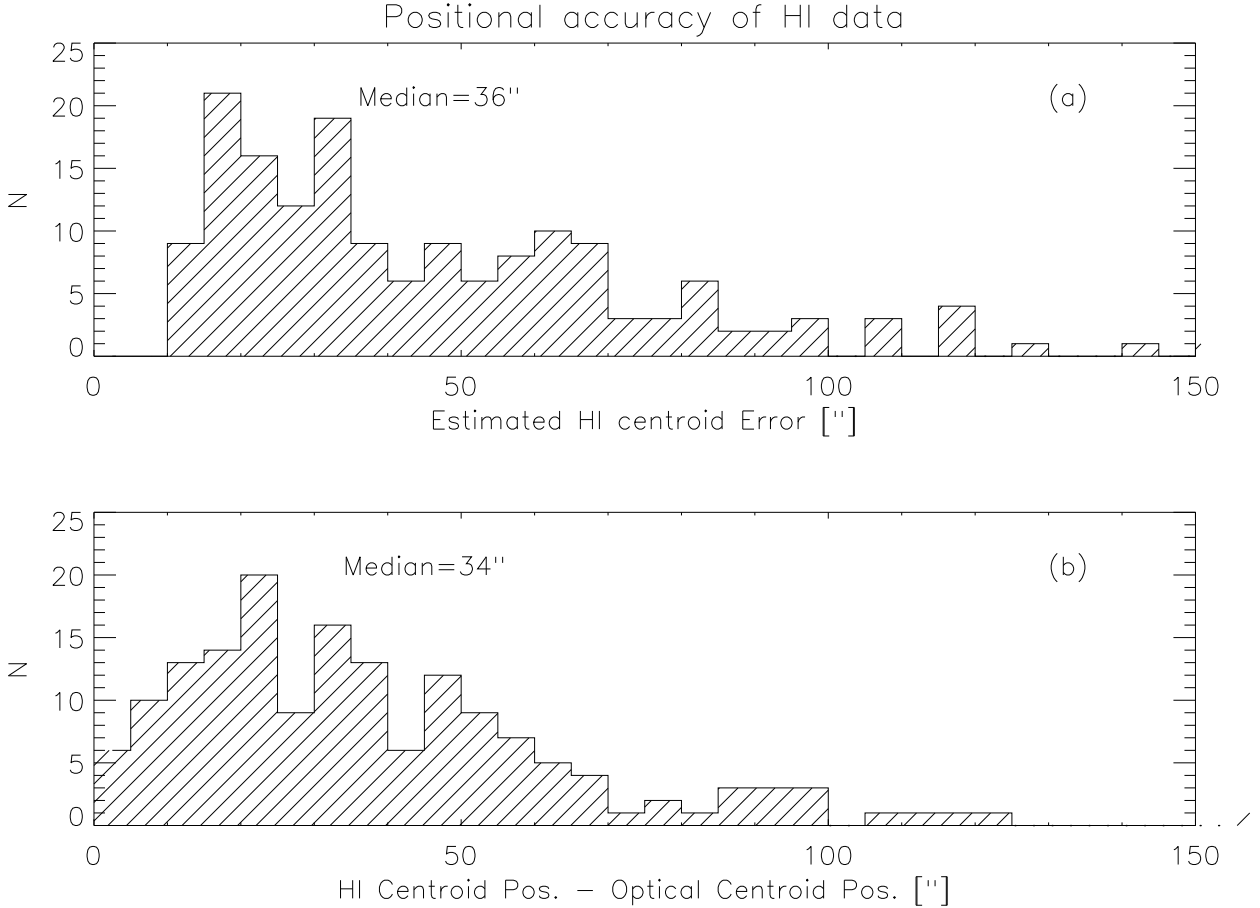}
\caption{(a) Histogram of the estimated uncertainty on the HI centroid position, 
$(\epsilon_\alpha^2+\epsilon_\delta^2)^{1/2}$, expressed in arcsec.
(b) Histogram of the positional difference between the HI centroid and the
center of the optical counterpart, in arcsec.}
\label{centroiderror}
\end{figure}

The accuracy of the positions of HI sources depends on (a) the pointing quality
of ALFA and (b) our ability to accurately centroid on maps of HI emission. While 
the pointing characteristics of ALFA are still being evaluated, the average 
pointing errors for all ALFA beams  --- as estimated from bright sources of
known positions --- are smaller than 15\arcsec. We
evaluate the map centroiding component simply
by propagating the uncertainties on flux integrals measured throughout
the source map, and combining it in quadrature with a 10\arcsec ~rms pointing error.
Estimated uncertainties on the HI centroids are tabulated in column 2 
of Table \ref{alfadet}. A histogram of the total HI centroid estimated errors is shown in
Figure \ref{centroiderror}(a). The median on the HI centroid total estimated 
positional uncertainty is 36\arcsec. In Figure \ref{centroiderror}(b), we plot the
histogram of the positional difference between the HI centroid and the optical
counterpart center. The median value of that difference is 34\arcsec, suggesting 
that our positional error estimates are statistically correct.

%FIGURE 2

\subsection{Velocities and Widths}

\begin{figure}[h]
%\plotone{./figs/cz_alfa_lbw.ps}
\plotone{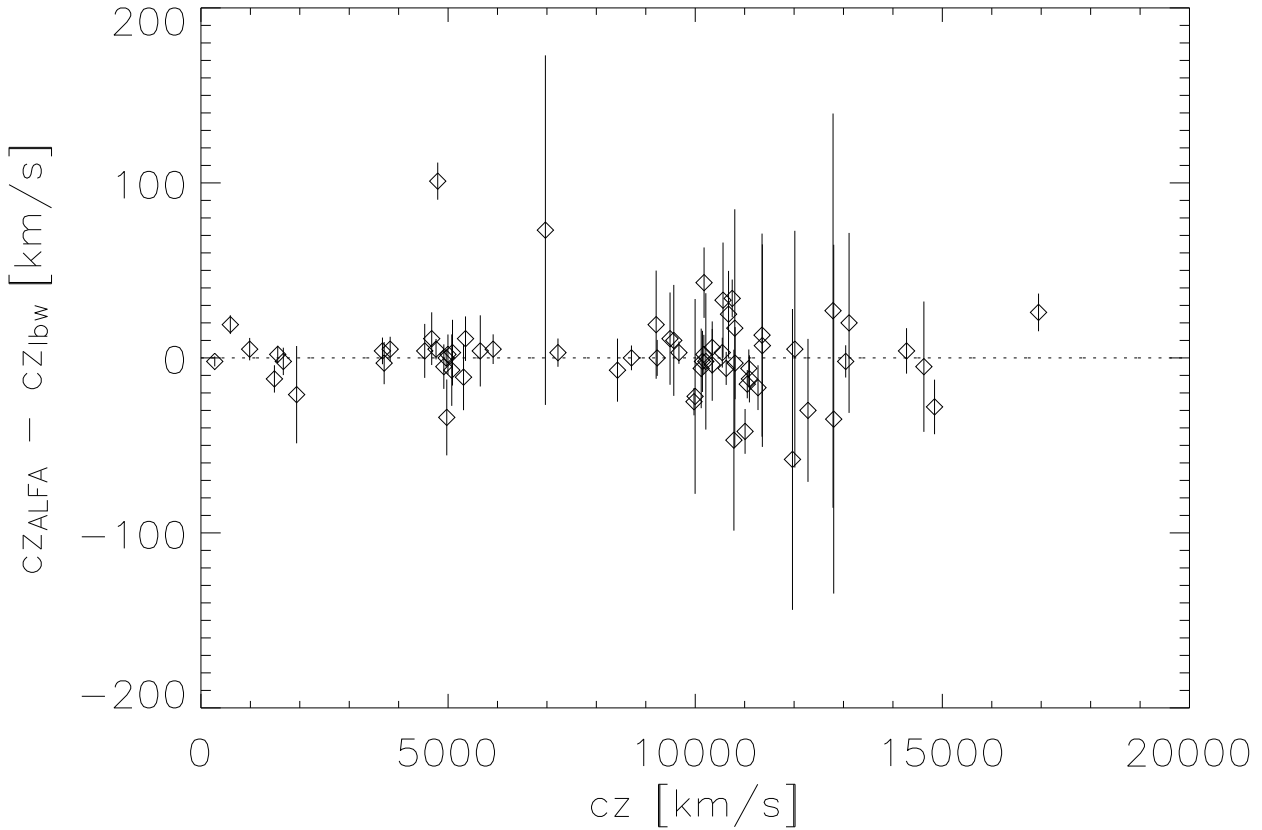}
\caption{Systemic velocity difference between the ALFA and LBW spectra for
68 objects observed with both systems. Note that these objects are among
the lowest signal--to--noise detections in the sample.}
\label{cz_alfa_lbw}
\end{figure}

\begin{figure}[h]
%\plotone{./figs/w_alfa_lbw.ps}
\plotone{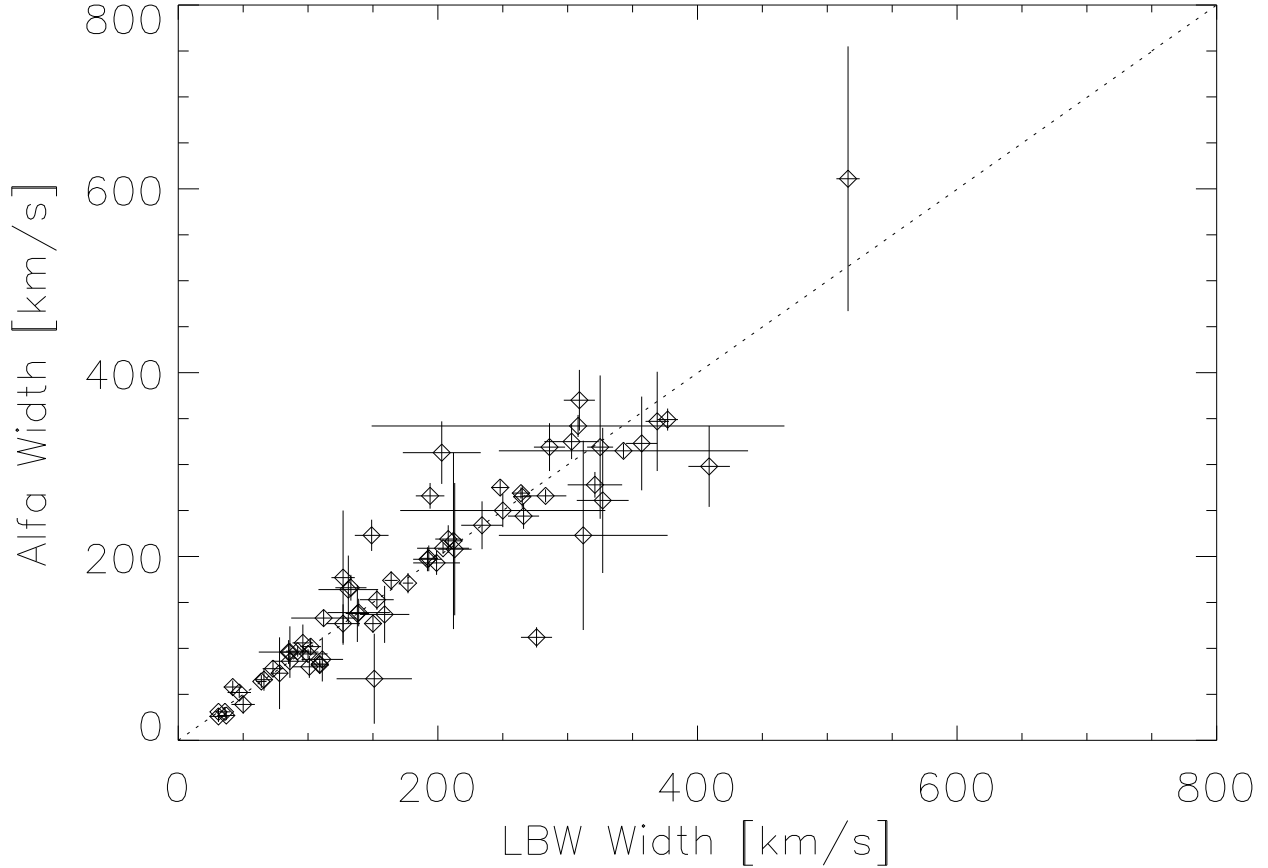}
\caption{Comparison of velocity widths as measured in the ALFA and LBW spectra for
68 objects observed with both systems. The diagonal line of slope 1 is
shown as a visual aid. Note that these objects are among
the lowest signal--to--noise detections in the sample.}
\label{w_alfa_lbw}
\end{figure}

Figure \ref{cz_alfa_lbw} displays the difference between the systemic velocities
measured respectively for the ALFA and for the LBW data, versus the systemic
velocity itself. Figure \ref{w_alfa_lbw} shows a comparison of the velocity widths 
measured on ALFA spectra with those measured on LBW spectra.
Apart from a couple of objects with clearly underestimated 
errors, no significant difference between the two scales is seen. Note that the
objects included in this comparison are among those with the lowest signal to noise
in our sample (which is the reason why their initial detection needed corroboration).

Springob \etal ~(2005) have constructed a digital archive of some 9000 
extragalactic HI spectra, the majority of which were obtained at Arecibo,
centered on optically-identified galaxies, and have reprocessed the digital
spectra to extract parameters using a homogeneous set of algorithms. Fifty--one of the
detections presented in Table \ref{alfadet} are also represented in the
optically-targeted HI archive; all of the archival spectra were obtained with 
one of the Arecibo dual polarization line feeds available before the Gregorian upgrade.
A comparison of the velocity widths measured on
ALFA spectra with those of previous HI observations given by Springob \etal,
allows a comparison for high signal to noise sources. The correlation between 
reported and newly measured widths is very good, guaranteeing the good quality 
of the extracted high signal to
noise ALFA data and its immediate applicability in studies which may involve
the luminosity--linewidth relation and other scaling relations.

%FIGURE 3  

%FIGURE 4  

%FIGURE 5  
%\begin{figure}[h]
%\plotone{./figs/w_pub_alfa.ps}
%\plotone{f5.eps}
%\caption{Comparison of velocity widths as measured on the ALFA spectra 
%with those for the same objects reported in the literature. The diagonal 
%line of slope 1 is shown as a visual aid. Note that these objects are 
%among the highest signal--to--noise detections in the sample.}
%\label{w_pub_alfa}
%\end{figure}

\subsection{HI Line Fluxes}

\begin{figure}[h]
%\figurenum{1}
%\plotone{./figs/fi_alfa_lbw.ps}
\plotone{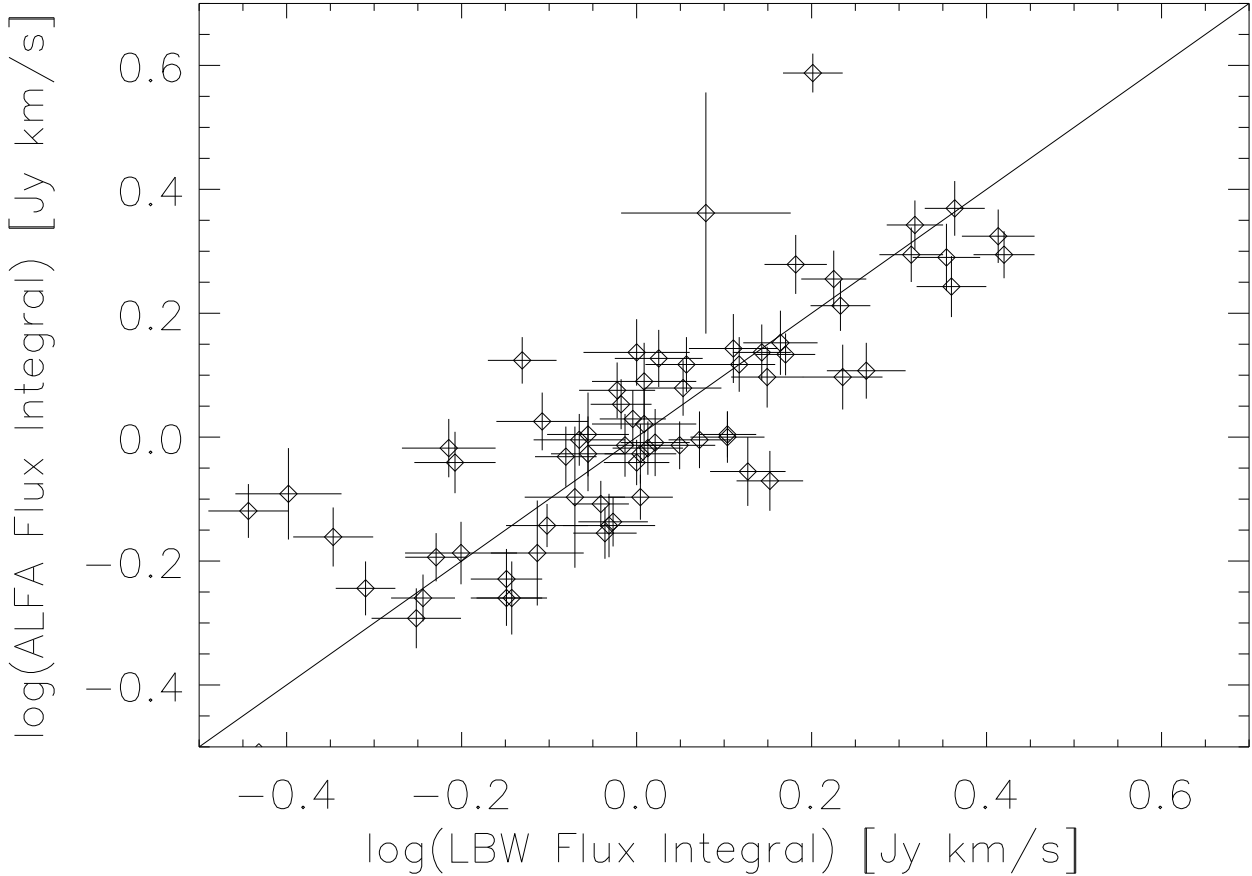}
\caption{Comparison of HI line flux integrals $F_c$ as measured in the ALFA 
and LBW spectra for 68 objects observed with both systems. The diagonal 
line of slope 1 is shown as a visual aid. Note that these objects 
are among the lowest signal--to--noise detections in the sample.}
\label{fi_alfa_lbw}
\end{figure}

\begin{figure}[h]
%\figurenum{1}
%\plotone{./figs/fi_pub_alfa.ps}
\plotone{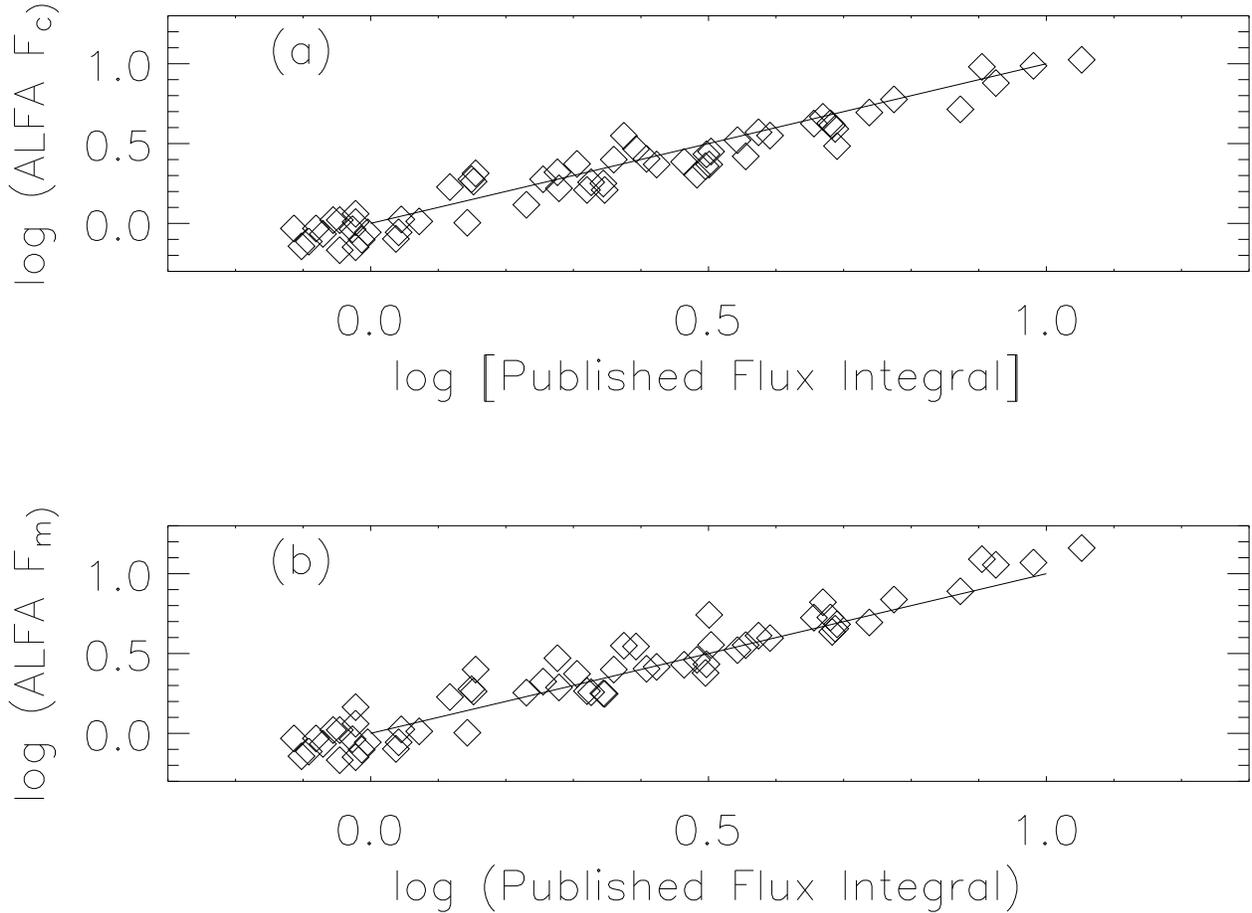}
\caption{(a) Comparison of HI flux integrals $F_c$, as measured in the ALFA 
composite spectra, with those available from the literature. $F_c$ are
uncorrected for beam dilution. (b) Comparison of HI line flux integrals $F_m$,
obtained by integrating ALFA maps of the sources, with those obtained from
the literature. The diagonal lines of slope 1 are shown as visual aids. }
\label{fi_pub_alfa}
\end{figure}

In Figure \ref{fi_alfa_lbw}, we compare the HI line flux integrals obtained from the
single beam LBW spectra with those obtained from the composite
spectra with ALFA. In spite of the large scatter, cross--calibration of the flux
density scales appears satisfactory. Again, note that this subset of the
data contains the objects closest to the detection limit of our survey.
Possibly a more relevant comparison is that obtained for the objects with
previously published HI flux integrals (Springob \etal ~2005). Figure \ref{fi_pub_alfa} shows
such a comparison. In panel (a), we compare published flux integrals with
those measured on the ALFA composite spectrum, which includes only the
flux seen by a single beam, centered on the HI centroid position.  The ALFA 
flux integrals fall slightly short of the published values, by about 15\%.
However, the published flux integrals, in their vast majority based on
Arecibo single beam observations, have been corrected by a statistical
dilution factor which takes into account the expected extent of the HI
emission, based on the optical size of the galaxy. In panel (b) of Figure 
\ref{fi_pub_alfa}, for those objects for which a significant indication
of extended HI emission is available, we use the ALFA flux integrals $F_m$,
which better represent the total flux integral. We exclude from the comparison
HI 014729.9+271958 (IC 1727) and HI 014753.9+272555 (NGC 672): the emission of 
the two galaxies is blended and a detailed analysis of the system is required
to produce accurate estimates of the respective fluxes, a task beyond the
scope of this preliminary report. Figure \ref{fi_pub_alfa} (a) and (b)
indicate that the recovery of HI flux integrals via ALFA is largely
consistent with well established previous efforts.

%FIGURE 6  

%FIGURE 7  

\subsection{Distances}

Source distances $D_{pec}$, as listed in Column 6 of Table \ref{optparms}, 
are obtained from the peculiar velocity model of the local Universe of Tonry 
\etal ~(2000). The model provides the expected peculiar velocity at a given 
point in space, and it can be inverted to yield the distance to a galaxy of 
given angular coordinates and observed redshift. We have used $D_{pec}$ to
compute the HI mass in Column 7 of Table \ref{optparms}. A word of caution
is however needed. The inversion of the peculiar velocity model yields 
single--valued solutions throughout the region populated by sources in Table 
\ref{optparms} (unlike, for example, of what would take place in the direction 
of the Virgo cluster). Throughout this region, the dipole component of the
peculiar velocity is negative. This component does not vanish at large
distances, and beyond 60 Mpc $D_{cmb}$ probably yields a more accurate
estimate of distance. Between $D_{cmb}=40$ to 60 Mpc, an average between
$D_{cmb}$ and $D_{pec}$ may be a better estimate of distance, although we note
that the differences between the two values are small. 

The adoption of such a peculiar velocity model yields, for nearby galaxies in 
the part of the sky sampled by this study, significantly larger estimates of 
distance than those obtained assuming a pure Hubble flow model, $D_{cmb}$. For 
example, for the 7 objects with $D_{cmb}<10$ Mpc, the mean $D_{pec}$ is about 6 Mpc 
larger than $D_{cmb}$. For those, the assumption of pure Hubble flow would yield 
distances which are smaller by more than a factor of 4 and HI masses that are smaller 
by more than a factor of 20, than if a peculiar velocity flow model is assumed. 
In addition to large--scale flows, the peculiar velocity model includes a `thermal'
component of 187 \kms, which translates in an additional distance uncertainty
of 2.7 Mpc. This illustrates how the assumption of a distance model can have an 
important effect in the derivation of the HI mass function, as its low mass slope 
is strongly dependent on the parameters of (few known) nearby objects
(Masters, Haynes \& Giovanelli 2004).

\section{Survey Sensitivity}\label{sensitivity}

While the sky coverage of the precursor observations was uneven and the sampling
density of some of the regions covered was poor (given the vagaries forced by the
telescope schedule), the data obtained lend themselves to a preliminary appraisal
of the sensitivity limits of the ALFALFA survey. As discussed in Paper I, a 2--pass 
drift survey, i.e. one with drift tracks spaced about 1\arcmin ~in Declination, yields an 
effective integration time per beam area of $\sim$48 sec, or a map with
effective integration per pixel area of 30 seconds, if no significant deterioration
of the spatial resolution is imposed. The signal extraction from the precursor
observations was however exercised on single beam drift tracks, which have an
rms of $3.5\times(res/10)^{-1/2}$ mJy, where $res$ is the spectral resolution 
of the data after smoothing, in \kms. The detection threshold of a given spectral 
feature depends on the width of the feature, as discussed in Paper I, so that 
spectral smoothing can be increased for a broad feature, and the rms noise 
accordingly decreased. This approach remains valid until a threshold width is 
reached, beyond which further smoothing is overcome by the effect of large--scale 
instabilities in the spectral baseline, which cannot be undone by smoothing. In
Paper I we assumed that the threshold width for our data lies in the vicinity of 
200 \kms. 

\begin{figure}[h]
%\figurenum{1}
%\plotone{./figs/w_fi_alfa.ps}
\plotone{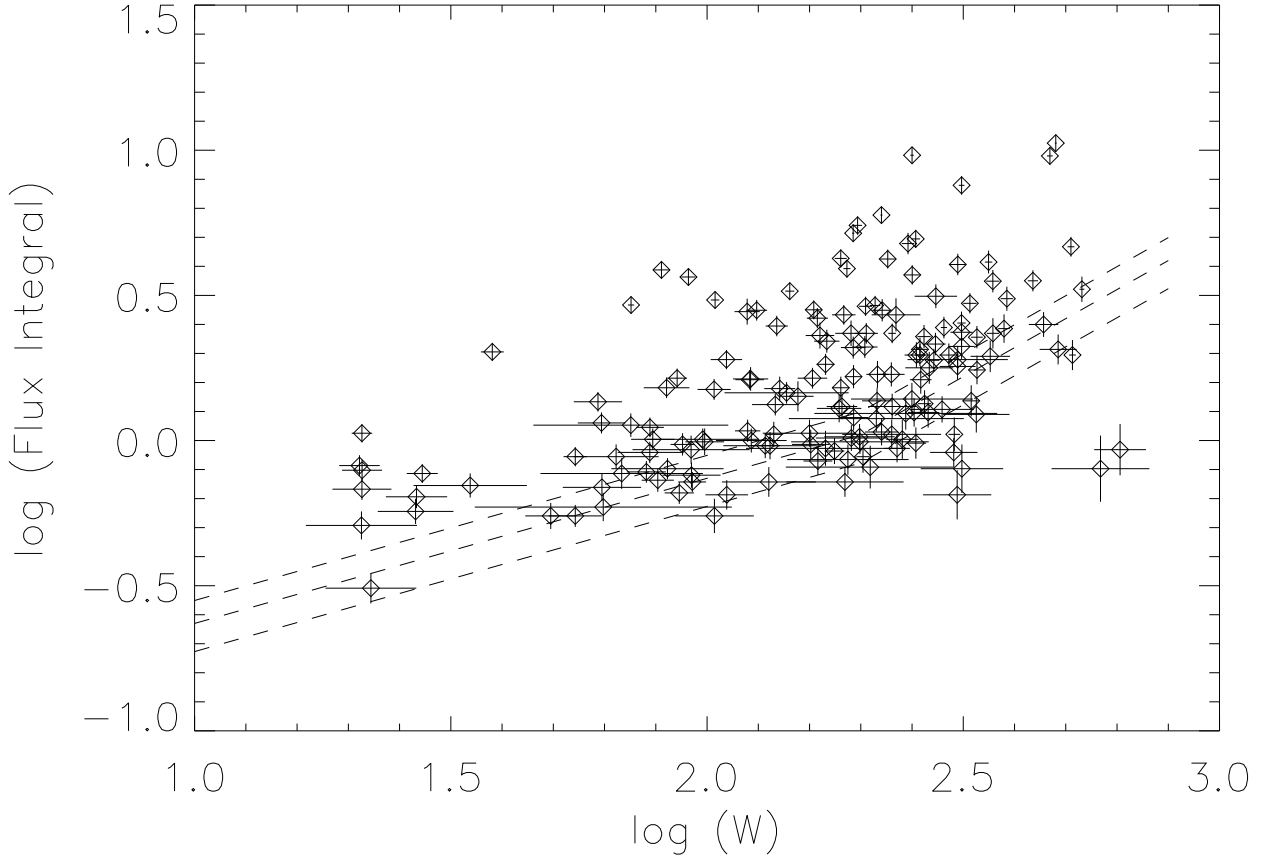}
\caption{HI flux integral plotted versus velocity width, for the detected sample.
The three dashed lines correspond to the flux threshold $F_{th}$ one obtains
for a signal--to--noise ratio of, respectively, 4, 5 and 6, using Eqn. \ref{eq.ston}.}
\label{w_fi_alfa}
\end{figure}

Figure \ref{w_fi_alfa} displays a plot of the HI line flux integral of the 
HI detections in our precursor run, versus their velocity widths. 
The figure corroborates the expectation that the sensitivity limit of the 
survey depends on velocity width, lower flux integrals being detectable at 
smaller widths. The dashed lines inset in the figure represent possible 
detection thresholds at different levels of signal--to--noise. 
Suppose the detection threshold is defined as
\be
F_{th}=(S/N)\times (rms)\times W
\ee
where $S/N$ is a signal--to--noise fiducial figure for detection and 
$rms$ is the root mean square noise over a pixel, after smoothing 
of the spectrum. With the noise rms expressed in Jy and the width 
in \kms, the flux integral $F_{th}$ is in Jy \kms. If the spectrum 
is smoothed to a resolution equal to $W/2$ for $W<200$ \kms, and to 
a resolution equal to $200/2$ \kms ~for $W\ge 200$ \kms, then for
the rms appropriate to single drift maps (cf. Paper I)

\be
%\scriptsize{
F_{th}= \left\{\begin{array}{ll}
           0.22~(S/N)~(W/200)^{1/2}  & \mbox{if $W<200$} \\
           0.22~(S/N)~(W/200)        & \mbox{if $W\ge 200$}
           \end{array}
           \right.
%}
\label{eq.ston}
\ee

The dashed lines in Figure \ref{w_fi_alfa} ~are Eqn. \ref{eq.ston} ~for 
nominal figures of $S/N$ 4, 5 and 6, which show that we are able to 
detect sources with flux well below the nominal sensitivity limit for 
a conservative signal--to--noise ratio of 6, as discussed in Paper I. 
A few sources of large
width are detected at signal--to--noise levels even significantly lower
than 4. This can be attributed to the fact that
extragalactic line profiles are not boxlike, but rather two--horned:
a spectral feature of $W=300$ \kms, which consists of two horns each 
35 \kms ~wide, can be detected with similar ease to that with which 
we can detect a 300 \kms ~boxlike spectral line with twice the flux 
integral. Figure \ref{w_fi_alfa} ~can be used to illustrate two
important points: 
\begin{itemize}
\item The spatially two--dimensional data cubes that will
be produced by ALFALFA will have a sensitivity per pixel 1.5 times
better than the single drift, position--velocity maps used to obtain 
candidate detections
in the precursor run; thus a conservative threshold of $S/N=6$ for
the ALFALFA survey is equivalent to the dashed line of $S/N=4$ in
Figure \ref{w_fi_alfa}. To that level, we expect the vast majority
of ALFALFA candidate detections will not need follow--up
corroborating observations.
\item ALFALFA candidate detections to $S/N\geq 4$ will deliver a 
substantial additional fraction of confirmations, with modest amount 
of follow--up telescope time, as discussed in Section 7 of Paper I.
\end{itemize}

\begin{figure}[h]
%\figurenum{1}
%\plotone{./figs/d_him.ps}
\plotone{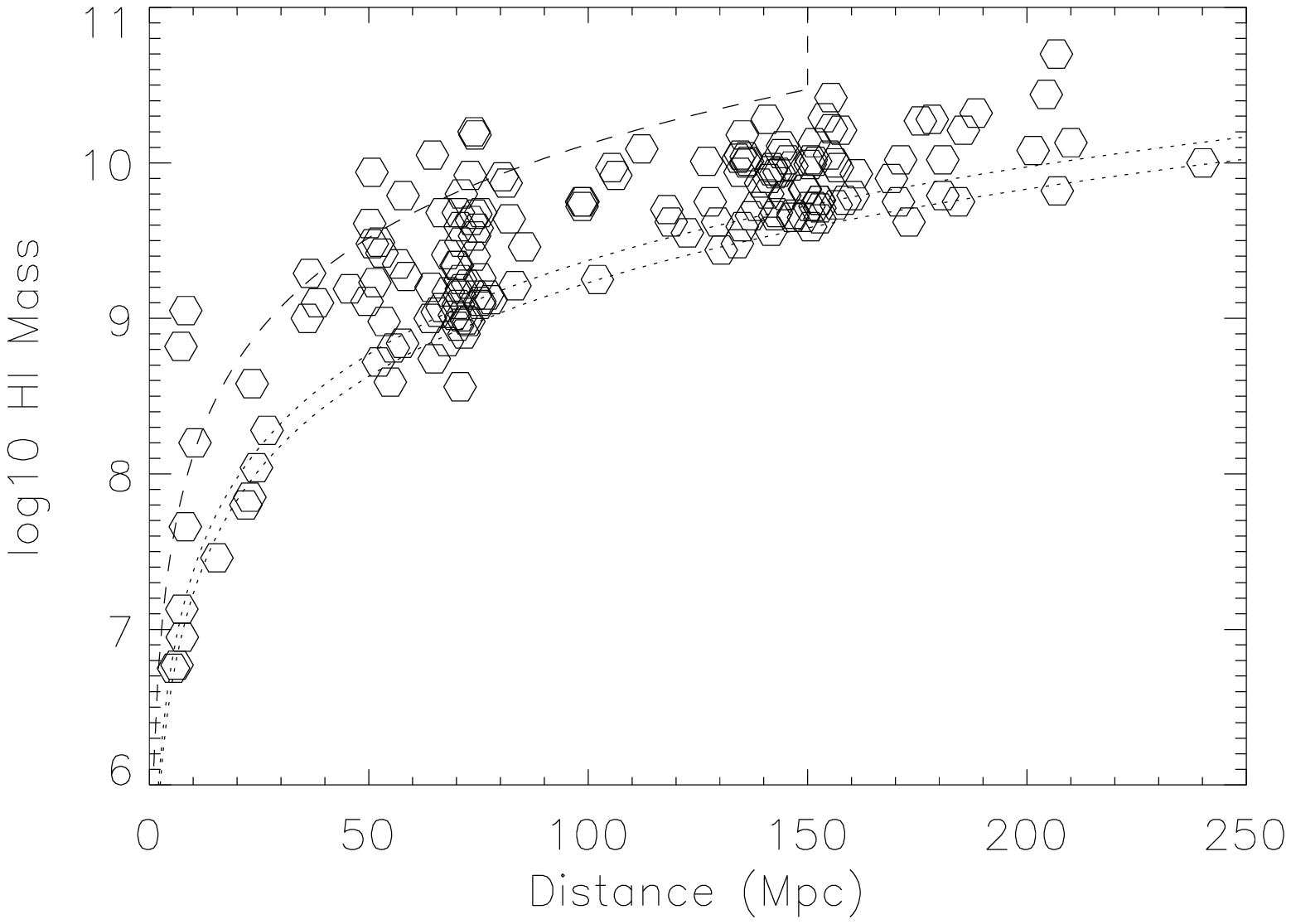}
\caption{Distribution of HI Mass as a function of distance, for the detected sample.
Distances are estimated from a peculiar velocity model of the Local Universe. The net 
effect of this model, for galaxies in this part of the sky, is that distance estimates 
are significantly larger than those that would be obtained from a pure Hubble flow
model. The highest of the two dotted lines corresponds to a flux integral of 1.0 
Jy \kms; the lower
dotted line corresponds to 0.72 Jy \kms, the 5--sigma limit for a $W=200$ \kms ~source
with an effective integration of 30 seconds per map pixel, which will apply to
the ALFALFA survey. The dashed line corresponds to a flux integral of 5.6 Jy \kms, 
which corresponds to the HIPASS $5\sigma$ detection limit for a
$W=200$ \kms ~source.}
\label{d_him}
\end{figure}

Figure \ref{d_him} shows the distribution of HI mass plotted as a 
function of source distance. The dotted lines in Figure \ref{d_him} 
correspond to  flux density integrals of 1. and 0.72 Jy \kms. The
lower level corresponds to the $5\sigma$ limit for a $W=200$ 
\kms ~source with an effective integration of 30 seconds per map pixel, 
which will apply to the ALFALFA survey. The dashed line corresponds 
to a flux density of 5.6 Jy \kms, the $5\sigma$ detection limit 
for a $W=200$ \kms ~source in the HIPASS survey, as obtained by the
same rationale as that by which our limits are estimated. It is 
apparent that, of the 166 objects detected by our experiment, 
HIPASS would have detected 10\% 
or fewer, and none of those farther than 100 Mpc.

%FIGURE 8

%FIGURE 9

%FIGURE 1  
%\begin{figure}[t]
%\figurenum{1}
% either psfig or epsscale+plotone will work below.
%\centerline{\psfig{figure=../rpe2_6.ps}}
%\epsscale{1.0}
%\plotone{fig1.ps}
%\caption{}
%\end{figure}

\section{Discussion}\label{discussion}

The call for precursor observations by NAIC during the period of ALFA 
commissioning was intended to provide potential ALFA survey teams with the opportunity 
to gain experience with the system on the earliest possible timescale, and to
contribute to the commissioning effort through debugging and testing of telescope
control and data taking hardware and software. From our perspective, this ``shared-risk''
collaboration has proved very advantageous: both the ALFALFA survey and the
ALFA system are better as a result.

While most of the data obtained during our precursor observations were not of science
quality, we nonetheless are greatly encouraged both by the detection
rate achieved in the precursor run and the
nature of the extragalactic HI sources presented here in Table
\ref{alfadet}. After less than 1\% of the planned full 
survey time and under less-than-optimal conditions, we can clearly
see the delivery of ALFALFA's science promise.

Even without a correction for incompleteness,
the redshift distribution of the detected sources carries clear signature of the
large scale structure in the distribution of galaxies in the Local Universe. 
Two strong peaks occur near 70 and 150 Mpc, as expected given that the region surveyed 
overlaps the Perseus--Pisces Supercluster. For the same reason, the number of 
sources nearer than 50 Mpc is much smaller than would be expected for a random sample,
because a prominently underdense void separates the Local Group from Pisces-Perseus. Because
of the extreme environmental conditions represented by the regions sampled, the
set of detected sources presented here does not lend itself to reliable inferences regarding
the HI mass function. As discussed in Paper I, the determination of the HI mass function
in a variety of environments is a main science objective of ALFALFA.

As illustrated in Figure \ref{d_him}, ALFALFA will survey the HI universe 
to a distance of 250 Mpc, allowing detection of massive HI rich objects.
The HI detection with the highest redshift found in the precursor
observations, HI 020329.8+273909, is identified with a very blue galaxy, 
found in the KISO survey with previously unknown redshift, near $cz = 17,000$ 
\kms, which places the HI line close to the edge of our bandpass.
The source with the highest HI mass, HI 020248.2+263434, is identified with 
the previously cataloged galaxy UGC~1533, a barred spiral seen nearly
face--on at a distance of 208 Mpc ($cz_\odot=14,600$ \kms). Its HI
mass, $5\times 10^{10}$ \msun, is about one order of magnitude
larger than that of the Milky Way. 

As discussed in detail in Paper I, 
for a HI survey conducted with a high sensitivity telescope like Arecibo, 
the probability of maximizing detection of the lowest mass objects is enhanced 
more by increasing the solid angle coverage than by increasing the survey depth. 
Indeed, these precursor observations have already detected three objects with 
HI masses lower than $10^7$ M$_\odot$, all located within 10 Mpc from us.
Given the model we have used for the peculiar velocity field, no object in our
detected sample was assigned a distance smaller than 5.5 Mpc. Note that the
``thermal'' rms peculiar velocity of galaxies translates into a distance 
uncertainty of about 2.7 Mpc, and thus the
distances of the nearest objects are highly uncertain. These objects belong to
the group of which the brightest members are NGC 672 (HI 014753.9+272555) and
IC1727 (HI 014729.9+271958). We detected 8 possible members of the group, namely:
HI 014105.8+272007, HI 014214.9+262202, HI 014441.4+271707, HI 014640.9+264754, 
HI 014729.9+271958, \
HI 014753.9+272555, HI 015519.2+275645, HI 021404.3+275302. 
Given the vicinity and abundance of gas--rich systems and the availability
of primary distances for several group members (7.9 Mpc for NGC 672, 5.4 Mpc
for UGC 1281, 4.7 Mpc for AGC111977=KK16 and for AGC111164=KK17 and 5.0 Mpc for
NGC 784), a detailed study of this group appears as potentially very rewarding
and will follow.

Albeit at very small distances, the detected objects with the lowest HI masses have 
small sizes, unresolved by the telescope beam, so that their diameters are less than
a few kpc. They are associated with very low surface brightness optical counterparts,
with optical diameters on the order of 1 kpc. The velocity widths of
these systems tend to be quite small, on the order of 25 \kms, and their halo masses
may well be less than $10^8$ \msun. The determination of the cosmic density of 
low mass, optically faint gas-rich dwarfs constitutes one of the main goals of ALFALFA.

The procedure for source identification which yielded the 166 detections presented here
was biased against narrow--line, small angular size sources. Because of the unusually
large amount of RFI with similar characteristics in these early ALFA data, we did not seek
detection confirmation of most narrow-width candidates. Since much of
the narrow RFI, especially at frequencies near 1420 MHz, was internally
generated and has since been eliminated, the prospects for detecting
narrow line HI signals has considerably improved since these early
ALFA observations. For that reason, and because the local volume probed by
these observations is significantly underdense relative to other
regions of the ALFALFA sky, we expect the overall detection rate of
low mass objects to be considerably higher than that of the current sample.
In fact, as discussed in greater detail in Paper I, we can
confidently predict that the full ALFALFA survey covering more than 7000 \sqd ~will
detect more than 100 times as many HI sources as reported here. We look forward to
the surprising objects among them.

\vskip 0.3in

RG and MPH acknowledge the partial support of NAIC as Visiting Scientists during
the period of this work. GLH acknowledges travel support from NAIC. 
This work has been supported by NSF grants AST--0307661, AST--0302049, 
AST--0435697, AST--0347929, AST--0407011; DGICT of Spain grant
AYA2003--07468--C03--01; RFBR grant 04-02-16115; and by a Brinson
Foundation grant. We thank Robert Brown, the Director of NAIC, for
stimulating the development of major ALFA surveys, Jeff Hagen and Mikael Lerner for 
numerous last minute fixes to CIMA, and Hector Hernandez for a sensible approach 
to telescope scheduling. 

%\newpage

%\input refs.tex

%\input paper2_table2.tex
%\input paper2_table3.tex
%\input paper2_table4.tex
\begin{deluxetable}{ccrrcrrl}
\tablewidth{0pt}
\tabletypesize{\scriptsize}
\tablecaption{Parameters of ALFA Detections\label{alfadet}}
\tablehead{
\colhead{Source}  & \colhead{$\epsilon_\alpha, \epsilon_\delta$}   &  
\colhead{cz$_\odot (\epsilon_{cz})$}  & \colhead{$W (\epsilon_w)$}  & \colhead{rms} &
\colhead{$F_{c} (\epsilon_{f})$} & \colhead{$F_m$} &  \colhead{Notes}    \\
{} &{sec,\,\,\,"} &{\kms} &{\kms} &{mJy} &
{Jy \kms} &{Jy \kms} &{}
}
\startdata
HI 001709.7+271616 &  4.2, 28 &   3707( 8, 0) &  150( 11) & 3.10 &  1.42(0.18) &       & L *\\
HI 002115.8+262318 &  2.3, 27 &   9210( 7,25) &  158( 37) & 2.02 &  1.06(0.12) &       & L *\\
HI 002312.5+272644 &  1.5, 20 &   3940( 8, 0) &  192( 11) & 2.30 &  2.09(0.19) &  2.96 &   *\\
HI 002818.8+272136 &  4.7, 62 &   9611( 8, 0) &  361( 11) & 2.27 &  3.54(0.33) &       &   *\\
HI 002905.0+272936 &  4.7, 25 &   8710( 5, 0) &   78(  8) & 2.00 &  1.01(0.09) &       & L *\\
HI 003027.1+263852 &  3.2, 31 &   9671( 4, 0) &  260(  6) & 1.81 &  1.97(0.18) &  2.49 & L  \\
HI 003052.6+272328 &  3.5, 43 &  11270(10, 0) &  159( 14) & 1.92 &  0.97(0.12) &       & L  \\
HI 003831.2+262132 &  1.3, 16 &   7223( 4, 0) &   93(  6) & 1.43 &  0.72(0.06) &       & L  \\
HI 003855.1+265753 &  0.8, 11 &   5178( 5, 0) &  193(  7) & 2.11 &  1.66(0.16) &  1.94 &   *\\
HI 004034.8+270239 &  0.9, 13 &   5134( 4, 0) &  124(  6) & 3.26 &  2.81(0.24) &  3.38 &   *\\
HI 004113.4+272248 &  2.4, 31 &  10344(14, 0) &  313( 19) & 1.71 &  2.11(0.22) &       & L  \\
HI 004256.8+271522 &  1.0, 13 &   5292( 7, 0) &  219( 10) & 2.63 &  2.81(0.26) &  3.57 &   *\\
HI 004357.1+260731 &  2.1, 57 &   4905( 1, 0) &   27(  2) & 1.90 &  0.77(0.06) &       &   *\\
HI 004404.2+261243 &  1.5, 18 &  10088( 6, 0) &  516(  8) & 1.65 &  1.97(0.25) &       &   *\\
HI 004411.5+265042 &  0.7, 10 &   5197( 1, 0) &  251(  2) & 1.81 &  9.61(0.68) & 11.73 &    \\
HI 004456.0+272758 &  1.6, 37 &   5205( 5, 0) &  119(  7) & 1.85 &  1.08(0.10) &       &   *\\
HI 004629.4+264134 &  2.3, 31 &   5089( 9, 0) &   93( 13) & 1.61 &  0.76(0.08) &       & L  \\
HI 004645.5+275553 &          &   6969( 3,84) &           & 1.77 &  2.30(1.30) &       & L *\\
HI 004650.7+262846 &  0.9, 13 &   4961( 3, 0) &  229(  4) & 1.63 &  2.34(0.19) &  2.61 &   *\\
HI 004716.1+274854 &  2.5, 32 &   4748(11, 0) &  313( 16) & 1.88 &  2.36(0.24) &       &   *\\
HI 004738.1+264856 &  1.1, 16 &   4956( 5, 0) &  103(  8) & 1.83 &  1.50(0.13) &       &    \\
HI 004803.3+273717 &  1.7, 23 &   4955( 5, 0) &  538(  6) & 1.99 &  3.32(0.35) &       &   *\\
HI 004816.9+262806 &  1.8, 23 &   5270( 6, 0) &  177(  8) & 1.63 &  0.92(0.10) &       &   *\\
HI 004834.5+274100 &  0.9, 19 &   5216( 4, 0) &  161(  6) & 2.00 &  2.82(0.22) &  3.58 &   *\\
HI 005009.5+271328 &  1.2, 16 &   4925( 4, 0) &  134(  5) & 2.07 &  1.05(0.11) &       &   *\\
HI 005124.7+271517 &  2.9, 39 &   4527( 9,11) &  103( 20) & 2.11 &  0.55(0.08) &       & L  \\
HI 005305.3+272909 &  1.2, 16 &   4756( 5, 0) &   61(  7) & 1.84 &  1.36(0.11) &       & L *\\
HI 010032.5+270540 &  2.1, 31 &  11008(10, 0) &  256( 14) & 2.36 &  1.97(0.21) &       & L  \\
HI 010158.0+263010 &  2.8, 38 &  10065(21, 0) &   68( 30) & 1.38 &  0.77(0.10) &       &   *\\
HI 010225.7+263713 &  1.6, 23 &   4973( 5,18) &  229( 26) & 1.74 &  1.31(0.14) &       & L *\\
HI 010318.1+264747 &  0.9, 12 &   5074(11, 0) &  135( 15) & 1.69 &  1.33(0.12) &       & L *\\
HI 010456.8+264942 &  1.0, 14 &  14275( 5, 0) &  253(  7) & 1.85 &  1.25(0.15) &       & L  \\
HI 010654.5+263054 &  1.8, 22 &  10147(15, 0) &  121( 21) & 1.56 &  1.00(0.10) &       & L  \\
HI 010703.4+271516 &  0.9, 12 &  10121( 9,20) &  132( 31) & 1.82 &  0.96(0.11) &       & L  \\
HI 011302.6+273832 &  1.6, 18 &   4961( 3, 0) &   22(  5) & 1.74 &  0.31(0.04) &       & L *\\
HI 011443.5+270813 &  0.8, 10 &   3625( 1, 0) &  103(  2) & 2.04 &  3.05(0.22) &  4.82 &   *\\
HI 011643.2+265307 &  3.7, 58 &  12282( 8,34) &   62( 49) & 1.83 &  0.59(0.07) &       & L  \\
HI 011743.9+270006 &  0.7, 11 &  12148( 7, 0) &  138( 10) & 2.36 &  1.51(0.15) &       &   *\\
HI 012242.5+265157 &  0.9, 14 &   9064( 7,32) &  142( 46) & 1.58 &  1.46(0.13) &       &   *\\
HI 012403.9+270300 &  0.6, 10 &   4923( 7, 0) &  156( 10) & 1.83 &  1.89(0.16) &       &   *\\
HI 012405.2+280433 &  0.9, 60 &   4084( 8, 0) &  166( 11) & 4.08 &  2.30(0.25) &       &   *\\
HI 012607.5+275810 &  1.3, 17 &   4031(10, 0) &   66( 14) & 1.91 &  0.88(0.09) &       &   *\\
HI 012944.1+272249 &  0.7, 10 &  12646( 8, 0) &  160( 11) & 1.70 &  1.64(0.14) &  2.55 &   *\\
HI 013152.5+271912 &  1.5, 22 &   3828( 5, 0) &   55(  7) & 1.75 &  0.55(0.05) &       & L  \\
HI 013315.1+261438 &  0.7, 10 &   6960( 3, 0) &  289(  4) & 1.90 &  2.45(0.22) &       &    \\
HI 013339.0+262439 &  1.1, 16 &  11096(10, 0) &  234( 14) & 1.27 &  0.94(0.11) &       & L  \\
HI 013716.0+262606 &  1.2, 17 &   3880( 2, 0) &   55(  3) & 1.62 &  0.88(0.07) &       &   *\\
HI 013740.2+271042 &  0.8, 11 &  11085( 9, 0) &  336( 12) & 1.68 &  1.75(0.21) &       & L  \\
HI 014105.8+272007 &  1.3, 18 &    280( 2, 0) &   27(  4) & 2.03 &  0.64(0.06) &       & L *\\
HI 014147.3+273159 &  5.3, 85 &  10830( 9,39) &  229( 57) & 1.69 &  1.05(0.13) &       &   *\\
HI 014214.9+262202 &  1.7, 23 &    364( 1, 0) &   21(  1) & 1.82 &  1.06(0.08) &       &   *\\
HI 014331.2+275930 &  3.2, 39 &  10179(18, 0) &  308( 26) & 1.67 &  1.80(0.20) &       & L  \\
HI 014428.3+275522 &  0.7,  9 &   4048( 1, 0) &  192(  1) & 1.85 &  5.18(0.38) &  7.76 &   *\\
HI 014441.4+271707 &  0.7, 10 &    430( 2, 0) &   38(  2) & 1.82 &  2.02(0.15) &  2.89 &   *\\
HI 014455.3+272942 &  4.3, 57 &  12802(10,99) &  585(144) & 1.55 &  0.80(0.24) &       & L *\\
HI 014640.9+264754 &  2.3, 31 &    370( 2, 0) &   21(  3) & 2.09 &  0.68(0.06) &       &   *\\
HI 014653.7+280448 &  2.0, 60 &   3597(19, 0) &  233( 27) & 3.90 &  2.71(0.38) &       &   *\\
HI 014724.3+275312 &  1.6, 22 &   3707( 7, 0) &  432( 10) & 1.76 &  3.55(0.31) &  3.95 &   *\\
HI 014729.9+271958 &          &    351( 2, 0) &  117(  3) & 1.88 & 54.39(3.81) &       &   *\\
HI 014753.9+272555 &          &    436( 2, 0) &  175(  3) & 1.77 & 69.25(4.85) &       &   *\\
HI 014837.9+273259 &  2.1, 28 &  11017( 8, 0) &  384( 11) & 2.10 &  3.08(0.31) &  2.92 &   *\\
HI 014915.3+274248 &  2.6, 32 &  10760( 9, 0) &  261( 13) & 1.91 &  1.62(0.19) &  1.90 &   *\\
HI 015011.6+271145 &  0.7, 10 &   3511( 1, 0) &  218(  1) & 1.69 &  5.98(0.43) &  6.89 &   *\\
HI 015013.0+273842 &  0.7,  9 &   3543( 1, 0) &  479(  2) & 1.81 & 10.58(0.77) & 14.47 &   *\\
HI 015047.7+271714 &  2.7, 34 &  14627( 9,35) &  307( 51) & 1.54 &  0.65(0.14) &       & L  \\
HI 015150.7+274448 &  1.7, 22 &  10749( 4, 0) &   89(  5) & 2.06 &  0.97(0.09) &       & L  \\
HI 015439.1+274824 &  1.2, 15 &  14842( 9, 0) &  183( 13) & 1.81 &  1.31(0.14) &       & L  \\
HI 015439.8+271111 &  1.0, 15 &   8428(10, 0) &  270( 14) & 1.75 &  1.25(0.16) &       & L *\\
HI 015519.2+275645 &  1.0, 13 &    219( 1, 0) &   21(  2) & 2.11 &  0.79(0.07) &       &   *\\
HI 015536.6+274154 &  2.3, 34 &   8356( 6, 0) &   83(  9) & 1.71 &  1.52(0.13) &       &    \\
HI 015751.8+264537 &  1.6, 22 &  12435( 9, 0) &  279( 13) & 1.83 &  2.15(0.20) &  2.54 &    \\
HI 015857.7+265710 &  1.9, 25 &  13044( 7, 0) &   34( 10) & 1.68 &  0.70(0.07) &       & L  \\
HI 015917.6+270027 &  1.9, 24 &   5313( 6, 0) &  129(  9) & 1.95 &  0.96(0.10) &       & L *\\
HI 015937.0+272555 &  1.2, 14 &   5276( 4, 0) &  203(  6) & 1.79 &  2.90(0.23) &       &   *\\
HI 015952.5+262407 &  1.0, 14 &   5226( 6, 0) &  109(  8) & 1.57 &  1.90(0.15) &  2.60 &   *\\
HI 020133.9+262914 &  0.8, 11 &   5155( 6, 0) &  512(  8) & 1.54 &  4.65(0.38) &  6.63 &   *\\
HI 020144.4+263227 &  0.8, 12 &   5014( 4, 0) &  182(  5) & 1.69 &  4.24(0.32) &  5.32 &   *\\
HI 020248.2+263434 &  0.8, 11 &  14603( 7, 0) &  204( 11) & 1.91 &  2.34(0.20) &  5.02 &   *\\
HI 020304.8+271222 &  3.0, 52 &   4790( 7, 0) &  109( 11) & 1.82 &  0.65(0.08) &       & L *\\
HI 020329.8+273909 &  2.9, 38 &  16946( 4, 0) &   80(  6) & 2.01 &  0.73(0.07) &       & L *\\
HI 020343.0+261608 &  2.5, 36 &   5009(11,40) &  188( 59) & 1.55 &  0.86(0.11) &       &   *\\
HI 020353.5+261719 &  1.9, 22 &   5060( 4, 0) &   88(  6) & 1.55 &  0.66(0.06) &       &   *\\
HI 020430.7+275454 &  0.8, 12 &   4710( 2, 0) &  187(  3) & 1.69 &  3.91(0.29) &  4.53 &   *\\
HI 020530.0+272806 &  5.5, 65 &  11968( 9,72) &  213(103) & 1.58 &  1.19(0.13) &       & L  \\
HI 020558.3+272121 &  1.5, 18 &   4939( 5, 0) &  182(  7) & 1.92 &  1.52(0.14) &  1.88 &    \\
HI 020626.5+270152 &  1.1, 19 &   4974( 6, 0) &  251(  9) & 1.86 &  3.72(0.30) &  4.09 &   *\\
HI 020902.1+273202 &  1.2, 19 &   9868( 6, 0) &  121(  8) & 1.60 &  1.62(0.13) &       &   *\\
HI 020913.9+264536 &  3.1, 41 &  12788( 9, 0) &  327( 12) & 1.95 &  1.37(0.18) &       & L *\\
HI 020954.1+273147 &  1.4, 23 &   4914( 9, 0) &   77( 12) & 1.61 &  0.91(0.08) &       & L *\\
HI 021309.0+274650 &  2.0, 29 &  13113( 8,50) &  198( 72) & 1.65 &  0.99(0.10) &  2.02 & L  \\
HI 021328.9+263102 &  1.4, 20 &  14436( 9, 0) &  296( 13) & 1.63 &  1.97(0.19) &  2.78 &    \\
HI 021404.3+275302 &  0.8, 12 &    594( 2, 0) &   81(  3) & 1.91 &  3.87(0.29) &  6.28 & L *\\
HI 021510.4+262415 &  1.9, 27 &   4998( 9, 0) &  193( 13) & 1.63 &  1.20(0.13) &       & L  \\
HI 021640.8+260835 &  3.7, 58 &  11352(13, 0) &  240( 18) & 2.20 &  1.01(0.17) &       & L  \\
HI 022102.1+274615 &  4.0, 53 &   9539( 5, 0) &  164(  8) & 2.11 &  2.64(0.22) &       &   *\\
HI 022103.9+270204 &  2.9, 38 &  10800( 6, 0) &  303(  8) & 1.74 &  1.05(0.15) &       & L *\\
HI 022138.2+280302 &  4.2, 48 &  10674( 8,22) &  356( 33) & 2.29 &  1.95(0.26) &       & L  \\
HI 022224.8+262552 &  1.4, 20 &  11143( 6, 0) &  336(  9) & 1.51 &  2.27(0.21) &  2.76 &   *\\
HI 022335.8+271851 &  3.8, 53 &  10662( 3,39) &  185( 56) & 1.59 &  0.72(0.09) &       &   *\\
HI 022340.2+270927 &  1.7, 24 &  10660( 5, 0) &  170(  7) & 1.67 &  1.83(0.16) &       &   *\\
HI 022345.8+262214 &  5.4, 83 &   9566( 9,22) &  302( 34) & 1.54 &  0.91(0.11) &       & L  \\
HI 022348.9+272848 &  7.0, 98 &  10718(10,19) &  453( 31) & 1.64 &  2.51(0.26) &       &   *\\
HI 022355.8+270618 &  2.3, 30 &   5474( 5,19) &   98( 28) & 1.85 &  1.01(0.09) &       &   *\\
HI 022405.6+263900 &  1.5, 21 &   9230( 9, 0) &   62( 12) & 2.22 &  0.69(0.08) &       & L *\\
HI 022459.4+260314 &  4.0, 75 &  10099( 4,22) &  271( 31) & 3.02 &  1.78(0.24) &       &   *\\
HI 022533.4+264458 &  2.7, 39 &  10363( 5, 0) &  277(  7) & 1.73 &  1.89(0.18) &       &   *\\
HI 022538.8+271709 &  1.1, 23 &   8997( 7, 0) &  184( 10) & 1.69 &  2.71(0.22) &       &   *\\
HI 022542.3+265853 &  3.2, 42 &   9976( 6, 0) &  265(  8) & 1.71 &  1.34(0.15) &       & L  \\
HI 022558.5+271607 &  1.0, 15 &  10215( 5,38) &  335( 54) & 1.96 &  1.23(0.19) &  2.65 & L *\\
HI 022609.4+273549 &  3.3, 46 &   9995(10, 0) &  228( 14) & 1.92 &  1.69(0.17) &       &   *\\
HI 022617.1+260750 &  4.1, 63 &  10250( 9,87) &  244(124) & 2.31 &  1.24(0.17) &       &   *\\
HI 022620.2+271315 &  5.5, 96 &  10384( 9,55) &  639( 79) & 1.52 &  0.93(0.21) &       &   *\\
HI 022629.9+273937 &  5.7, 63 &   9717( 7,22) &  191( 33) & 1.59 &  1.02(0.11) &       &   *\\
HI 022632.1+274941 &  3.4, 52 &   9632( 6, 0) &  265(  8) & 2.00 &  2.28(0.22) &       &   *\\
HI 022740.9+265524 &  2.9, 43 &  10548( 8, 0) &  164( 11) & 1.82 &  0.85(0.10) &       & L  \\
HI 022741.1+271328 &  2.5, 47 &   5125(12,45) &  314( 64) & 1.69 &  0.80(0.17) &       &   *\\
HI 022742.7+261406 &  3.5, 62 &   9505( 5, 0) &  313(  6) & 1.85 &  2.54(0.24) &       &   *\\
HI 022745.2+263507 &  1.6, 24 &   9806( 5, 0) &   62(  7) & 1.84 &  1.15(0.10) &       &   *\\
HI 022751.5+275429 &  2.7, 34 &  10562( 8,30) &  287( 44) & 1.55 &  1.28(0.14) &       & L *\\
HI 022816.3+261854 &  0.7, 11 &   5249( 4, 0) &  466(  6) & 1.65 &  9.56(0.70) & 12.33 &   *\\
HI 022850.0+264500 &  1.6, 23 &   9529( 6, 0) &  260(  9) & 1.99 &  2.06(0.20) &       &    \\
HI 022900.2+263041 &  4.2, 68 &   9490(10,20) &  131( 31) & 1.75 &  0.72(0.09) &       & L  \\
HI 022901.5+260509 &  1.7, 20 &   5912( 6, 0) &   98(  9) & 2.65 &  0.99(0.11) &       & L  \\
HI 022919.3+272107 &  1.7, 25 &   1553( 4, 0) &   26(  5) & 2.43 &  0.57(0.06) &       & L  \\
HI 023023.1+275653 &  1.6, 25 &   4746( 4, 0) &   77(  5) & 2.00 &  1.11(0.09) &       &    \\
HI 023052.0+261047 &  2.5, 40 &  13300(10,28) &  484( 42) & 1.62 &  2.06(0.26) &  2.51 &   *\\
HI 023058.9+275709 &  0.8, 11 &   1596( 1, 0) &   71(  1) & 2.13 &  2.93(0.21) &       &    \\
HI 023103.8+274053 &  3.9, 75 &   4549( 5,24) &  198( 34) & 1.59 &  1.03(0.11) &       &   *\\
HI 023121.5+261239 &  3.1, 49 &   2523( 2, 0) &  145(  3) & 1.87 &  3.27(0.24) &       &    \\
HI 023124.6+264706 &  1.0, 14 &  10979( 7, 0) &  325( 10) & 1.63 &  2.97(0.25) &  4.67 &    \\
HI 023133.0+264727 &  0.8, 12 &   4559( 3, 0) &   87(  4) & 1.71 &  1.64(0.13) &       &    \\
HI 023135.9+263155 &  1.5, 23 &   3701( 1, 0) &   20(  2) & 1.94 &  0.82(0.07) &       &    \\
HI 023137.5+261010 &  1.1, 15 &  10874( 4, 0) &  212(  5) & 1.89 &  2.92(0.23) &  3.50 &   *\\
HI 023138.9+271130 &  2.6, 45 &    985( 4, 0) &   21(  6) & 2.87 &  0.51(0.06) &       & L  \\
HI 023203.1+261156 &  1.8, 25 &  10807( 4,17) &   83( 24) & 1.77 &  0.80(0.07) &       & L  \\
HI 023234.8+275608 &  3.0, 37 &   4668( 5,11) &  219( 17) & 1.73 &  1.07(0.12) &       & L  \\
HI 023325.2+271204 &  3.2, 44 &   5352( 7, 8) &  214( 15) & 2.04 &  1.37(0.15) &       & L  \\
HI 023329.2+270136 &  2.2, 35 &   1666( 5, 0) &   76(  7) & 1.90 &  0.78(0.07) &       & L  \\
HI 023516.3+280108 &  0.7, 12 &   2705( 2, 0) &   91(  3) & 2.67 &  3.66(0.27) &       &    \\
HI 023851.0+275109 &  0.7, 10 &   4595( 3, 0) &  313(  4) & 2.71 &  7.57(0.58) & 11.37 &   *\\
HI 024055.2+264046 &  4.0, 75 &   1487( 5, 0) &   49(  6) & 2.30 &  0.55(0.06) &       & L  \\
HI 024124.0+262227 &  3.1, 68 &  10627( 8, 0) &  255( 12) & 1.73 &  0.98(0.13) &       & L  \\
HI 024416.4+260648 &  2.4, 39 &  10571( 7,11) &  181( 19) & 2.36 &  1.29(0.15) &       &   *\\
HI 024600.5+280145 &  1.5, 26 &   7959( 4, 0) &  354(  6) & 2.84 &  4.12(0.39) &       &   *\\
HI 024609.7+270247 &  1.0, 14 &   5732( 6, 0) &  225(  8) & 2.50 &  4.22(0.33) &  5.30 &   *\\
HI 024752.7+270607 & 10.0, 15 &   5845( 5, 0) &  119(  7) & 6.47 &  2.78(0.30) &       &   *\\
HI 025122.4+263459 &  1.3, 21 &   7489( 5, 0) &  136(  7) & 3.06 &  2.48(0.21) &  3.57 &   *\\
HI 025149.8+260303 &  2.9, 59 &  10175(10, 0) &  190( 14) & 3.12 &  2.34(0.25) &       & L  \\
HI 030942.0+262649 &  2.3, 55 &  10783( 6,51) &  170( 73) & 2.54 &  0.94(0.14) &       & L  \\
HI 031952.6+262935 &  2.0, 29 &  11053( 7, 0) &  121( 10) & 2.91 &  1.63(0.16) &       & L  \\
HI 032120.9+262104 &  3.0, 47 &  11358( 4,56) &  251( 79) & 2.55 &  1.39(0.19) &       & L  \\
HI 033140.7+260603 &  3.1, 43 &  10344( 5, 0) &  201(  6) & 2.12 &  0.88(0.12) &       & L  \\
HI 033401.6+261019 &  4.9, 79 &   9997( 5,55) &  308( 78) & 2.22 &  1.90(0.22) &       & L  \\
HI 033716.0+262357 &  5.9, 80 &  12015(14,66) &  208( 96) & 2.11 &  0.81(0.15) &       & L *\\
HI 040226.1+264950 &  0.8, 16 &   5652( 6,19) &   93( 28) & 2.72 &  0.93(0.11) &       & L *\\
HI 040328.6+262149 &  2.6, 30 &   7072( 7, 0) &  379( 10) & 2.43 &  2.43(0.29) &       &   *\\
HI 041559.8+265855 &  1.8, 20 &   3673( 7, 0) &  171( 10) & 2.85 &  2.20(0.21) &       & L  \\
HI 041844.5+265513 &  1.9, 28 &   7582(12,16) &  279( 28) & 2.62 &  3.14(0.31) &       &    \\
HI 041904.2+261210 &  2.9, 44 &   3760( 4, 0) &  255(  5) & 2.07 &  4.95(0.37) &       &   *\\
HI 042739.5+260545 &  4.4, 96 &   1936( 9,26) &   70( 39) & 3.28 &  1.13(0.11) &       & L *\\
HI 060125.5+260524 &  1.1, 75 &   5934( 5, 0) &  246(  7) & 3.71 &  4.77(0.42) &       &   *\\
HI 060628.3+262314 &  0.7, 11 &   2724( 3, 0) &  196(  4) & 2.81 &  5.51(0.42) &  6.20 &   *\\
HI 062218.4+262631 &  2.7, 48 &   6196( 8, 0) &  215( 11) & 2.49 &  1.69(0.19) &       &   *\\
HI 063337.5+262911 &  1.3, 19 &   3404( 4, 8) &  203( 12) & 2.37 &  2.10(0.19) &  3.13 &    \\
HI 063840.8+263006 &  2.4, 32 &   9821( 8, 0) &  308( 11) & 2.82 &  4.04(0.36) &       &   *\\
HI 065004.2+262342 &  4.0, 56 &   9508( 7, 0) &  361( 10) & 2.67 &  2.34(0.30) &       &   *\\
\hline
\enddata
\end{deluxetable}

\begin{deluxetable}{cccccccc}
\tablewidth{0pt}
\tabletypesize{\scriptsize}
\tablecaption{Parameters of LBW Detections\label{LBWdet}}
\tablehead{
\colhead{Source}  & \colhead{R. A.}   &  \colhead{Dec.} &
\colhead{cz$_\odot (\epsilon_{cz})$}  & \colhead{$W (\epsilon_w)$}  & \colhead{rms} &
\colhead{$F_{c} (\epsilon_{f})$} & \colhead{Notes}    \\
{} &{hhmmss.s} & {ddmmss} &{\kms} &{\kms} &{mJy} &
{Jy \kms} &{}
}
\startdata
HI 001709.7+271616 &   001713.3 &  +271454 &   3710( 9, 0) &  150( 13) &  2.08 &  1.46(0.15) & *  \\
HI 002115.8+262318 &   002115.3 &  +262307 &   9191( 5,16) &  126( 23) &  2.27 &  0.78(0.10) & *  \\
HI 002905.0+272936 &   002905.0 &  +272914 &   8710( 5, 0) &  104(  7) &  1.42 &  1.27(0.10) & *  \\
HI 003027.1+263852 &   003026.8 &  +263833 &   9668( 5, 0) &  255(  7) &  1.84 &  2.63(0.22) &   \\
HI 003052.6+272328 &   003051.4 &  +272324 &  11287( 8, 0) &  127( 12) &  1.28 &  0.97(0.10) &   \\
HI 003831.2+262132 &   003832.3 &  +262050 &   7220( 7, 0) &   88(  9) &  2.10 &  0.79(0.09) &   \\
HI 004113.4+272248 &   004113.0 &  +272220 &  10348(15, 0) &  292( 21) &  1.95 &  2.59(0.26) &   \\
HI 004629.4+264134 &   004627.7 &  +264131 &   5086( 4,16) &   82( 23) &  1.27 &  0.36(0.04) &   \\
HI 004645.5+275553 &   004647.1 &  +275634 &   6896(10,53) &  289( 76) &  1.14 &  1.20(0.30) & *  \\
HI 005124.7+271517 &   005123.4 &  +271617 &   4523( 6, 0) &   93(  8) &  1.41 &  0.72(0.07) &   \\
HI 005305.3+272909 &   005307.8 &  +272909 &   4751( 3, 0) &   61(  4) &  1.75 &  1.48(0.12) & *  \\
HI 010032.5+270540 &   010034.0 &  +270553 &  11050( 8, 0) &  186( 11) &  1.50 &  2.06(0.18) &   \\
HI 010225.7+263613 &   010224.0 &  +263719 &   5007(11, 0) &  229( 16) &  1.31 &  1.14(0.13) & *  \\
HI 010318.1+264747 &   010317.8 &  +264747 &   5081(10,14) &  135( 24) &  1.12 &  0.74(0.07) & *  \\
HI 010456.8+264942 &   010456.5 &  +264942 &  14271(12, 0) &  269( 16) &  1.21 &  1.41(0.14) &   \\
HI 010654.5+263054 &   010700.8 &  +263054 &  10149( 9, 0) &  121( 13) &  2.14 &  1.27(0.13) &   \\
HI 010703.4+271516 &   010703.2 &  +271516 &  10127( 6, 0) &  132(  9) &  1.87 &  0.61(0.08) &   \\
HI 011302.6+273832 &   011257.5 &  +273758 &   4961( 4, 0) &   33(  6) &  1.00 &  0.37(0.03) &  * \\
HI 011643.2+265307 &   011643.0 &  +265445 &  12312( 7,20) &  144( 29) &  1.36 &  0.71(0.07) &   \\
HI 013152.5+271912 &   013154.0 &  +271944 &   3823( 5, 0) &   38(  7) &  1.44 &  0.57(0.05) &   \\
HI 013339.0+262439 &   013339.2 &  +262513 &  11108( 9, 0) &  256( 12) &  1.02 &  1.01(0.10) &   \\
HI 013740.2+271042 &   013743.7 &  +271050 &  11091( 6, 0) &  363(  8) &  1.60 &  2.29(0.22) &   \\
HI 014105.8+272007 &   014108.0 &  +271920 &    282( 2, 0) &   27(  3) &  1.61 &  0.59(0.05) &  * \\
HI 014331.2+275930 &   014327.0 &  +280002 &  10136( 9, 0) &  276( 12) &  1.24 &  1.68(0.15) &   \\
HI 014455.3+272942 &   014454.1 &  +272909 &  12837( 6, 0) &  494(  9) &  0.86 &  0.85(0.12) & *  \\
HI 015047.7+271714 &   015046.6 &  +271641 &  14632( 9, 0) &  340( 12) &  0.86 &  0.77(0.10) &   \\
HI 015150.7+274448 &   015152.7 &  +274531 &  10715(10, 0) &   95( 15) &  1.85 &  1.12(0.11) &   \\
HI 015439.1+274824 &   015438.9 &  +274919 &  14870( 9, 9) &  189( 18) &  1.70 &  1.31(0.13) &   \\
HI 015439.8+271111 &   015440.0 &  +271110 &   8435(15, 0) &  311( 21) &  1.61 &  1.72(0.19) & *  \\
HI 015857.7+265710 &   015859.4 &  +265749 &  13046( 6, 0) &   45(  9) &  1.94 &  0.92(0.08) &   \\
HI 015917.6+270027 &   015919.2 &  +270044 &   5324( 8,16) &  109( 25) &  1.96 &  1.03(0.10) & *  \\
HI 020304.8+271222 &   020303.0 &  +271316 &   4689( 8, 0) &  271( 12) &  1.21 &  0.63(0.10) &   \\
HI 020329.8+273909 &   020328.8 &  +273860 &  16920(10, 0) &   80( 14) &  1.77 &  0.94(0.09) &   \\
HI 020530.0+272806 &   020533.3 &  +272755 &  12026(10,45) &  299( 65) &  0.94 &  0.95(0.10) &   \\
HI 020913.9+264536 &   020918.5 &  +264528 &  12761( 8,112)&  295(159) &  1.65 &  1.00(0.15) & *  \\
HI 020954.1+273147 &   020956.3 &  +273234 &   4919( 9, 0) &   98( 13) &  1.44 &  1.00(0.09) &   \\
HI 021309.0+274650 &   021307.3 &  +274653 &  13093( 9, 0) &  203( 13) &  1.60 &  0.86(0.11) &   \\
HI 021404.3+275302 &   021405.8 &  +275034 &    575( 5, 0) &  107(  6) &  1.58 &  1.59(0.13) & *  \\
HI 021510.4+262415 &   021508.1 &  +262313 &   4996( 7, 0) &  188( 11) &  1.63 &  1.13(0.12) &   \\
HI 021640.8+260835 &   021639.7 &  +260816 &  11339( 8,56) &  240( 79) &  1.33 &  0.88(0.10) &   \\
HI 022103.9+270204 &   022101.0 &  +270204 &  10783(15,66) &  330( 96) &  1.56 &  1.02(0.15) & * \\
HI 022138.2+280302 &   022139.7 &  +280307 &  10649( 8, 0) &  298( 12) &  1.69 &  2.26(0.21) &   \\
HI 022345.8+262214 &   022347.5 &  +262207 &   9556(21, 0) &  196( 30) &  0.81 &  0.62(0.07) &   \\
HI 022405.6+263900 &   022406.5 &  +263960 &   9230( 5, 0) &   62(  7) &  1.65 &  0.45(0.05) & *  \\
HI 022542.3+265853 &   022545.2 &  +265824 &  10001( 5, 0) &  239(  7) &  1.74 &  1.06(0.13) &   \\
HI 022558.5+271607 &   022558.7 &  +271613 &  10217( 7, 0) &  356(  9) &  1.47 &  1.02(0.15) & *  \\
HI 022740.9+265524 &   022738.5 &  +265445 &  10545( 3, 0) &  170(  4) &  1.76 &  1.42(0.13) &   \\
HI 022751.5+275429 &   022754.1 &  +275419 &  10529(11, 0) &  394( 16) &  1.40 &  1.83(0.20) & *  \\
HI 022900.2+263041 &   022859.4 &  +263242 &   9479(14, 0) &  153( 19) &  1.92 &  0.93(0.12) &   \\
HI 022901.5+260509 &   022901.5 &  +260448 &   5907( 6, 0) &   98(  8) &  1.78 &  1.18(0.10) &   \\
HI 022919.3+272107 &   022919.9 &  +272116 &   1551( 3, 0) &   32(  4) &  1.11 &  0.49(0.04) &   \\
HI 023138.9+271130 &   023139.5 &  +271046 &    980( 5, 0) &   27(  7) &  2.54 &  0.56(0.07) &   \\
HI 023203.1+261156 &   023207.6 &  +261140 &  10810(11, 0) &  106( 16) &  1.63 &  1.01(0.09) &   \\
HI 023234.8+275608 &   023233.7 &  +275628 &   4657( 9, 0) &  145( 13) &  1.38 &  0.99(0.09) &   \\
HI 023325.2+271204 &   023328.5 &  +271139 &   5341( 7, 0) &  203( 10) &  1.03 &  1.39(0.11) &   \\
HI 023329.2+270136 &   023332.3 &  +270113 &   1668( 6, 0) &   71(  8) &  1.28 &  0.91(0.07) &   \\
HI 024055.2+264046 &   024055.6 &  +264005 &   1499( 6, 0) &   44(  9) &  1.69 &  0.71(0.07) &   \\
HI 024124.0+262227 &   024124.5 &  +262321 &  10633( 5, 0) &  255(  7) &  0.96 &  1.05(0.10) &   \\
HI 025149.8+260303 &   025150.0 &  +260310 &  10173( 5, 0) &  186(  7) &  1.62 &  2.31(0.19) &   \\
HI 030942.0+262649 &   030942.3 &  +262613 &  10830( 6, 0) &  121(  9) &  1.57 &  0.88(0.09) &   \\
HI 031952.6+262935 &   031955.0 &  +262945 &  11068( 4, 0) &  143(  6) &  1.66 &  1.71(0.14) &   \\
HI 032120.9+262104 &   032121.6 &  +262034 &  11351(14, 0) &  314( 20) &  1.57 &  1.29(0.16) &   \\
HI 033140.7+260603 &   033141.6 &  +260515 &  10338(14, 0) &  196( 20) &  1.53 &  1.34(0.14) &   \\
HI 033401.6+261019 &   033401.5 &  +261033 &  10019( 7, 0) &  314( 10) &  0.94 &  1.52(0.13) &   \\
HI 033716.0+262357 &   033722.5 &  +262505 &  12010( 5, 0) &  203(  7) &  0.91 &  0.40(0.06) & *  \\
HI 040226.1+264950 &   040226.3 &  +264928 &   5648( 4, 0) &   83(  6) &  1.59 &  0.83(0.07) & *  \\
HI 041559.8+265855 &   041600.0 &  +265934 &   3669( 3, 0) &  161(  5) &  1.56 &  2.08(0.16) &   \\
HI 042739.5+260545 &   042741.8 &  +260345 &   1957( 4, 0) &   76(  6) &  1.23 &  0.96(0.08) & *  \\
\enddata
\end{deluxetable}

\begin{deluxetable}{crccrrrc}
\tablewidth{0pt}
\tabletypesize{\scriptsize}
\tablecaption{Optical ID, Distances and HI Masses\label{optparms}}
\tablehead{
\colhead{Source}  & \colhead{AGC} & \colhead{R. A.}   &  \colhead{Dec.} &
\colhead{$D_{cmb}$}  & \colhead{$D_{pec}$}  & \colhead{log$(M_{HI})$} & \colhead{Notes} \\
{} & {} & {hhmmss.s} & {ddmmss} & Mpc & Mpc & $M_\odot$ & 
}
\startdata
HI 001709.7+271616 &   101815 &   001713.3 &  +271454 &  48.2 &  53.5 &  8.98 & * \\
HI 002115.8+262318 &   101816 &   002115.3 &  +262307 & 126.8 & 129.5 &  9.62 & * \\
HI 002312.5+272644 &      221 &   002310.9 &  +272555 &  51.6 &  56.8 &  9.35 & * \\
HI 002818.8+272136 &      278 &   002819.2 &  +272208 & 132.6 & 135.1 & 10.18 & * \\
HI 002905.0+272936 &   101817 &   002905.1 &  +272913 & 119.7 & 122.6 &  9.55 & * \\
HI 003027.1+263852 &   101819 &   003026.8 &  +263833 & 133.5 & 136.0 & 10.04 &   \\
HI 003052.6+272328 &   101820 &   003051.4 &  +272324 & 156.3 & 158.5 &  9.76 &   \\
HI 003831.2+262132 &   101821 &   003832.3 &  +262050 &  98.5 & 102.1 &  9.25 &   \\
HI 003855.1+265753 &   100381 &   003855.7 &  +265758 &  69.3 &  74.0 &  9.40 & * \\
HI 004034.8+270239 &   101822 &   004036.4 &  +270302 &  68.7 &  73.4 &  9.63 & * \\
HI 004113.4+272248 &   101823 &   004113.0 &  +272220 & 143.2 & 145.5 & 10.02 &   \\
HI 004256.8+271522 &   100482 &   004257.9 &  +271512 &  71.0 &  75.5 &  9.68 & * \\
HI 004357.1+260731 &   100489 &   004357.9 &  +260849 &  65.5 &  70.2 &  8.95 & * \\
HI 004404.2+261243 &      469 &   004403.7 &  +261227 & 139.5 & 141.9 &  9.97 & * \\
HI 004411.5+265042 &      470 &   004412.5 &  +265045 &  69.7 &  74.2 & 10.18 &   \\
HI 004456.0+272758 &   100499 &   004456.7 &  +272656 &  69.8 &  74.3 &  9.15 & * \\
HI 004629.4+264134 &   101825 &   004627.7 &  +264131 &  68.1 &  72.8 &  8.98 &   \\
HI 004645.5+275553 &   100550 &   004647.0 &  +275630 &  95.0 &  98.6 &  9.72 & * \\
HI 004650.7+262846 &      483 &   004650.3 &  +262831 &  66.3 &  71.0 &  9.49 & * \\
HI 004716.1+274854 &      489 &   004718.6 &  +274938 &  63.3 &  68.0 &  9.41 & * \\
HI 004738.1+264856 &   101826 &   004737.7 &  +264819 &  66.2 &  70.9 &  9.25 &   \\
HI 004803.3+273717 &      491 &   004801.5 &  +273728 &  66.2 &  70.8 &  9.59 & * \\
HI 004816.9+262806 &   100573 &   004816.6 &  +262802 &  70.7 &  75.3 &  9.09 & * \\
HI 004834.5+274100 &      497 &   004834.9 &  +274130 &  70.0 &  74.4 &  9.67 & * \\
HI 005009.5+271328 &   101827 &   005012.2 &  +271325 &  65.8 &  70.4 &  9.09 & * \\
HI 005124.7+271517 &   101828 &   005124.7 &  +271601 &  60.2 &  64.9 &  8.74 &   \\
HI 005305.3+272909 &   101829 &   005307.8 &  +272909 &  63.4 &  68.1 &  9.17 & * \\
HI 010032.5+270540 &   101685 &   010034.0 &  +270553 & 152.8 & 155.0 & 10.05 &   \\
HI 010158.0+263010 &   100714 &   010158.4 &  +262912 & 139.3 & 141.7 &  9.56 & * \\
HI 010225.7+263713 &   100719 &   010224.0 &  +263717 &  66.6 &  71.1 &  9.19 & * \\
HI 010318.1+264747 &   112506 &   010317.8 &  +264747 &  68.1 &  72.5 &  9.22 & * \\
HI 010456.8+264942 &   112507 &   010456.5 &  +264942 & 199.5 & 201.4 & 10.08 &   \\
HI 010654.5+263054 &   112508 &   010700.8 &  +263054 & 140.6 & 142.9 &  9.68 &   \\
HI 010703.4+271516 &   112509 &   010703.2 &  +271516 & 140.2 & 142.5 &  9.66 &   \\
HI 011302.6+273832 &   112510 &   011257.4 &  +273759 &  66.6 &  70.8 &  8.56 & * \\
HI 011443.5+270813 &   110150 &   011445.6 &  +270809 &  47.5 &  52.2 &  9.49 & * \\
HI 011643.2+265307 &   112511 &   011643.0 &  +265445 & 171.2 & 173.1 &  9.62 &   \\
HI 011743.9+270006 &   112512 &   011742.4 &  +270033 & 169.3 & 171.2 & 10.02 & * \\
HI 012242.5+265157 &   110263 &   012241.8 &  +265203 & 125.3 & 127.8 &  9.75 & * \\
HI 012403.9+270300 &      948 &   012404.1 &  +270246 &  66.1 &  70.2 &  9.34 & * \\
HI 012405.2+280433 &   112513 &   012403.6 &  +280557 &  54.2 &  58.5 &  9.27 & * \\
HI 012607.5+275810 &   110324 &   012608.8 &  +275756 &  53.4 &  57.7 &  8.84 & * \\
HI 012944.1+272249 &   112515 &            &          & 176.5 & 178.4 & 10.28 & * \\
HI 013152.5+271912 &   112516 &   013154.0 &  +271945 &  50.6 &  54.9 &  8.59 &   \\
HI 013315.1+261438 &   112518 &   013315.6 &  +261455 &  95.3 &  98.5 &  9.75 &   \\
HI 013339.0+262439 &   112519 &   013339.2 &  +262513 & 154.4 & 156.5 &  9.73 &   \\
HI 013716.0+262606 &   110443 &   013715.3 &  +262611 &  51.4 &  55.6 &  8.81 & * \\
HI 013740.2+271042 &   112520 &   013743.7 &  +271040 & 154.3 & 156.3 & 10.00 &   \\
HI 014105.8+272007 &   112521 &   014108.0 &  +271920 &       &   6.3 &  6.77 & * \\
HI 014147.3+273159 &   110477 &   014146.4 &  +273012 & 150.7 & 152.7 &  9.76 & * \\
HI 014214.9+262202 &   110482 &   014217.3 &  +262200 &   1.2 &   7.4 &  7.13 & * \\
HI 014331.2+275930 &   111466 &   014327.0 &  +280004 & 141.5 & 143.5 &  9.94 &   \\
HI 014428.3+275522 &   110790 &   014428.8 &  +275544 &  53.9 &  57.8 &  9.79 & * \\
HI 014441.4+271707 &   111945 &   014442.7 &  +271718 &   2.2 &   8.2 &  7.66 & * \\
HI 014455.3+272942 &   112522 &   014454.1 &  +272909 & 178.9 & 180.8 &  9.79 & * \\
HI 014640.9+264754 &   111946 &   014642.2 &  +264805 &   1.3 &   7.5 &  6.95 & * \\
HI 014653.7+280448 &   111332 &   014654.2 &  +280558 &  47.5 &  51.4 &  9.23 & * \\
HI 014724.3+275312 &     1250 &   014725.0 &  +275309 &  49.0 &  53.0 &  9.42 & * \\
HI 014729.9+271958 &     1249 &   014729.9 &  +271958 &   1.1 &   7.2 &  8.82 & * \\
HI 014753.9+272555 &     1256 &   014753.9 &  +272555 &   2.3 &   8.3 &  9.05 & * \\
HI 014837.9+273259 &   110535 &   014835.2 &  +273253 & 153.5 & 155.4 & 10.22 & * \\
HI 014915.3+274248 &   110543 &   014920.1 &  +274244 & 149.8 & 151.8 & 10.01 & * \\
HI 015011.6+271145 &     1291 &   015012.3 &  +271145 &  46.3 &  50.2 &  9.61 & * \\
HI 015013.0+273842 &     1292 &   015014.0 &  +273845 &  46.7 &  50.7 &  9.94 & * \\
HI 015047.7+271714 &   112523 &   015046.6 &  +271641 & 205.1 & 206.9 &  9.82 &   \\
HI 015150.7+274448 &   112524 &   015152.7 &  +274531 & 149.7 & 151.7 &  9.72 &   \\
HI 015439.1+274824 &   112525 &   015438.9 &  +274919 & 208.2 & 210.0 & 10.13 &   \\
HI 015439.8+271111 &   112526 &   015440.0 &  +271110 & 116.6 & 118.9 &  9.62 & * \\
HI 015519.2+275645 &   111977 &   015520.4 &  +275713 &       &   5.5 &  6.75 & * \\
HI 015536.6+274154 &   112527 &   015535.6 &  +274328 & 115.6 & 117.9 &  9.70 &   \\
HI 015751.8+264537 &   112528 &   015752.2 &  +264543 & 173.9 & 175.7 & 10.27 &   \\
HI 015857.7+265710 &   112529 &   015859.6 &  +265749 & 182.6 & 184.4 &  9.75 &   \\
HI 015917.6+270027 &   112459 &   015919.2 &  +270044 &  72.1 &  75.3 &  9.11 & * \\
HI 015937.0+272555 &   110716 &   015938.1 &  +272601 &  71.6 &  74.7 &  9.58 & * \\
HI 015952.5+262407 &   112530 &   015953.0 &  +262416 &  70.9 &  74.1 &  9.53 & * \\
HI 020133.9+262914 &     1507 &   020130.8 &  +262851 &  69.9 &  73.1 &  9.92 & * \\
HI 020144.4+263227 &     1510 &   020146.3 &  +263246 &  67.9 &  71.1 &  9.80 & * \\
HI 020248.2+263434 &     1533 &   020248.1 &  +263452 & 204.9 & 206.7 & 10.70 & * \\
HI 020304.8+271222 &   122180 &            &          &  64.7 &  67.9 &  8.85 & * \\
HI 020329.8+273909 &   122181 &   020328.8 &  +273900 & 238.4 & 240.1 & 10.00 & * \\
HI 020343.0+261608 &   120005 &   020341.0 &  +261636 &  67.8 &  71.0 &  9.01 & * \\
HI 020353.5+261719 &   122182 &   020358.3 &  +261611 &  68.6 &  71.7 &  8.90 & * \\
HI 020430.7+275454 &     1565 &   020432.0 &  +275531 &  63.6 &  66.7 &  9.68 & * \\
HI 020530.0+272806 &   122183 &   020533.3 &  +272755 & 167.3 & 169.1 &  9.90 &   \\
HI 020558.3+272121 &   122185 &   020600.8 &  +272128 &  66.9 &  69.9 &  9.34 &   \\
HI 020626.5+270152 &     1595 &   020627.3 &  +270159 &  67.4 &  70.4 &  9.68 & * \\
HI 020902.1+273202 &   120055 &   020902.4 &  +273210 & 137.3 & 139.3 &  9.87 & * \\
HI 020913.9+264536 &   122186 &   020918.5 &  +264528 & 179.1 & 180.8 & 10.02 & * \\
HI 020954.1+273147 &   122187 &   020956.3 &  +273234 &  66.6 &  69.5 &  9.02 & * \\
HI 021309.0+274650 &   122188 &   021307.3 &  +274653 & 183.8 & 185.5 & 10.21 &   \\
HI 021328.9+263102 &   122189 &   021331.2 &  +263049 & 202.6 & 204.4 & 10.44 &   \\
HI 021404.3+275302 &     1718 &   021403.6 &  +275236 &   4.9 &  10.4 &  8.20 & * \\
HI 021510.4+262415 &   122190 &   021508.1 &  +262313 &  67.8 &  70.7 &  9.15 &   \\
HI 021640.8+260835 &   122191 &   021639.7 &  +260816 & 158.6 & 160.4 &  9.79 &   \\
HI 022102.1+274615 &   122192 &   022103.4 &  +274526 & 132.8 & 134.7 & 10.05 & * \\
HI 022103.9+270204 &   122193 &   022101.1 &  +270204 & 150.8 & 152.6 &  9.76 & * \\
HI 022138.2+280302 &   122194 &   022139.8 &  +280307 & 149.1 & 150.8 & 10.02 &   \\
HI 022224.8+262552 &   122195 &   022226.1 &  +262610 & 155.7 & 157.5 & 10.21 & * \\
HI 022335.8+271851 &   120201 &   022331.3 &  +271924 & 148.9 & 150.7 &  9.59 & * \\
HI 022340.2+270927 &     1844 &   022339.4 &  +270932 & 148.9 & 150.6 &  9.99 & * \\
HI 022345.8+262214 &   122196 &   022347.5 &  +262207 & 133.2 & 135.1 &  9.59 &   \\
HI 022348.9+272848 &     1848 &   022355.5 &  +272932 & 149.7 & 151.5 & 10.13 & * \\
HI 022355.8+270618 &     1850 &   022358.5 &  +270700 &  74.8 &  77.1 &  9.15 & * \\
HI 022405.6+263900 &   122197 &            &          & 128.4 & 130.3 &  9.44 & * \\
HI 022459.4+260314 &   120218 &   022500.2 &  +260301 & 140.9 & 142.7 &  9.93 & * \\
HI 022533.4+264458 &     1881 &   022535.2 &  +264436 & 144.6 & 146.4 &  9.98 & * \\
HI 022538.8+271709 &   120234 &   022539.5 &  +271743 & 125.1 & 127.0 & 10.01 & * \\
HI 022542.3+265853 &   122198 &   022545.2 &  +265824 & 139.1 & 140.9 &  9.80 &   \\
HI 022558.5+271607 &   121216 &   022558.7 &  +271613 & 142.6 & 144.3 & 10.11 & * \\
HI 022609.4+273549 &     1892 &   022612.5 &  +273610 & 139.4 & 141.2 &  9.90 & * \\
HI 022617.1+260750 &   120244 &   022615.2 &  +260623 & 143.0 & 144.8 &  9.79 & * \\
HI 022620.2+271315 &     1894 &   022616.8 &  +271242 & 145.0 & 146.7 &  9.67 & * \\
HI 022629.9+273937 &     1899 &   022626.2 &  +273913 & 135.4 & 137.2 &  9.66 & * \\
HI 022632.1+274941 &     1902 &   022633.7 &  +274926 & 134.2 & 136.0 & 10.00 & * \\
HI 022740.9+265524 &   122199 &   022738.5 &  +265445 & 147.3 & 149.1 &  9.65 &   \\
HI 022741.1+271328 &   120265 &   022740.9 &  +271315 &  69.9 &  72.2 &  8.99 & * \\
HI 022742.7+261406 &   120266 &   022742.3 &  +261334 & 132.4 & 134.2 & 10.03 & * \\
HI 022745.2+263507 &     1921 &   022746.4 &  +263520 & 136.7 & 138.5 &  9.72 & * \\
HI 022751.5+275429 &   121307 &   022754.9 &  +275422 & 147.5 & 149.3 &  9.83 & * \\
HI 022816.3+261854 &     1939 &   022817.6 &  +261844 &  71.6 &  74.0 & 10.20 & * \\
HI 022850.0+264500 &   122200 &   022849.0 &  +264450 & 132.8 & 134.6 &  9.94 &   \\
HI 022900.2+263041 &   122201 &   022859.4 &  +263242 & 132.2 & 134.0 &  9.48 &   \\
HI 022901.5+260509 &   122202 &   022901.5 &  +260448 &  81.1 &  83.3 &  9.21 &   \\
HI 022919.3+272107 &   122203 &   022919.9 &  +272116 &  18.9 &  22.9 &  7.85 &   \\
HI 023023.1+275653 &   122204 &   023024.1 &  +275736 &  64.5 &  66.7 &  9.07 &   \\
HI 023052.0+261047 &   120299 &   023050.4 &  +261136 & 186.7 & 188.4 & 10.32 & * \\
HI 023058.9+275709 &   122206 &   023100.3 &  +275730 &  19.5 &  23.4 &  8.58 &   \\
HI 023103.8+274053 &     1990 &   023104.1 &  +274158 &  61.7 &  64.0 &  9.00 & * \\
HI 023121.5+261239 &   122207 &   023122.1 &  +261152 &  32.7 &  36.0 &  9.00 &   \\
HI 023124.6+264706 &   122208 &   023124.5 &  +264717 & 153.5 & 155.3 & 10.42 &   \\
HI 023133.0+264727 &   122210 &   023133.0 &  +264752 &  61.8 &  64.2 &  9.20 &   \\
HI 023135.9+263155 &   122211 &   023136.8 &  +263230 &  49.6 &  52.2 &  8.72 &   \\
HI 023137.5+261010 &   120307 &   023134.6 &  +260958 & 152.0 & 153.8 & 10.29 & * \\
HI 023138.9+271130 &   122212 &   023139.3 &  +271045 &  10.8 &  15.4 &  7.46 &   \\
HI 023203.1+261156 &   122213 &   023207.6 &  +261140 & 151.1 & 152.8 &  9.64 &   \\
HI 023234.8+275608 &   122214 &   023233.7 &  +275628 &  63.4 &  65.6 &  9.04 &   \\
HI 023325.2+271204 &   122215 &   023328.5 &  +271139 &  73.2 &  75.3 &  9.26 &   \\
HI 023329.2+270136 &   122217 &   023332.3 &  +270113 &  20.5 &  24.4 &  8.04 &   \\
HI 023516.3+280108 &   122218 &   023515.2 &  +280116 &  35.4 &  38.3 &  9.10 &   \\
HI 023851.0+275109 &     2134 &   023851.9 &  +275048 &  62.5 &  64.5 & 10.05 & * \\
HI 024055.2+264046 &   122219 &   024055.6 &  +264005 &  18.1 &  22.0 &  7.80 &   \\
HI 024124.0+262227 &   122220 &   024124.5 &  +262321 & 148.7 & 150.3 &  9.72 &   \\
HI 024416.4+260648 &   122221 &   024418.5 &  +260629 & 147.9 & 149.5 &  9.83 & * \\
HI 024600.5+280145 &   120490 &   024601.1 &  +280140 & 110.6 & 112.2 & 10.09 & * \\
HI 024609.7+270247 &     2236 &   024611.1 &  +270239 &  78.8 &  80.4 &  9.91 & * \\
HI 024752.7+270607 &     2272 &   024804.5 &  +270609 &  80.5 &  82.0 &  9.64 & * \\
HI 025122.4+263459 &   122223 &   025123.1 &  +263458 & 104.0 & 105.4 &  9.97 & * \\
HI 025149.8+260303 &   122224 &   025150.0 &  +260310 & 142.4 & 143.9 & 10.06 &   \\
HI 030942.0+262649 &   131043 &   030942.3 &  +262613 & 151.4 & 152.8 &  9.71 &   \\
HI 031952.6+262935 &   131044 &   031955.5 &  +262936 & 155.4 & 156.8 &  9.97 &   \\
HI 032120.9+262104 &   131045 &   032121.6 &  +262034 & 159.8 & 161.1 &  9.93 &   \\
HI 033140.7+260603 &   131046 &   033141.6 &  +260515 & 145.5 & 146.6 &  9.65 &   \\
HI 033401.6+261019 &   131047 &   033401.5 &  +261033 & 140.6 & 141.7 &  9.95 &   \\
HI 033716.0+262357 &   131048 &   033722.5 &  +262505 & 169.5 & 170.7 &  9.75 & * \\
HI 040226.1+264950 &   130368 &   040226.3 &  +264928 &  79.1 &  77.9 &  9.12 & * \\
HI 040328.6+262149 &   140002 &   040329.8 &  +262146 &  99.4 &  98.8 &  9.75 & * \\
HI 041559.8+265855 &   140472 &   041559.7 &  +265955 &  51.0 &  49.6 &  9.11 &   \\
HI 041844.5+265513 &   140473 &   041843.8 &  +265526 & 106.9 & 106.3 &  9.92 &   \\
HI 041904.2+261210 &     3009 &   041905.9 &  +261048 &  52.3 &  50.8 &  9.48 & * \\
HI 042739.5+260545 &   140474 &   042743.5 &  +260437 &  26.5 &  26.8 &  8.28 & * \\
HI 060125.5+260524 &   150264 &   060126.6 &  +260328 &  85.5 &  81.3 &  9.87 & * \\
HI 060628.3+262314 &   160446 &   060628.9 &  +262346 &  39.7 &  36.5 &  9.29 & * \\
HI 062218.4+262631 &   160544 &            &          &  89.6 &  85.5 &  9.46 & * \\
HI 063337.5+262911 &   160545 &   063339.0 &  +262905 &  50.0 &  45.6 &  9.19 &   \\
HI 063840.8+263006 &   160122 &   063838.1 &  +263022 & 141.7 & 140.8 & 10.28 & * \\
HI 065004.2+262342 &   160242 &   065004.2 &  +262248 & 137.5 & 136.3 & 10.01 & * \\
\enddata
\end{deluxetable}

\end{document}